%% file: main_arxiv.tex
\documentclass[conference,compsoc]{IEEEtran}

\ifCLASSOPTIONcompsoc
  \usepackage[nocompress]{cite}
\else
  \usepackage{cite}
\fi

\usepackage{tikz}
\usepackage{filecontents}
\usepackage{xspace} 

\usepackage{amsmath,amssymb,amsfonts}
\usepackage{graphicx}
\usepackage{textcomp}
\usepackage{xcolor}
\usepackage{todonotes}
\usepackage{mathtools}
\usepackage{algorithm}
\usepackage[noend]{algpseudocode}
\usepackage[caption=false]{subfig}
\usepackage{tabularx}
\usepackage{booktabs}
\usepackage{makecell}
\usepackage{multirow}
\usepackage{comment}
\usepackage{url}

\makeatletter
\newcommand{\linebreakand}{%
  \end{@IEEEauthorhalign}
  \hfill\mbox{}\par
  \mbox{}\hfill\begin{@IEEEauthorhalign}
}
\makeatother

\ifCLASSINFOpdf
\else
\fi

\input{macros}

\begin{document}
\title{NeST: Neuron Selective Tuning for LLM Safety}

\author{
\IEEEauthorblockN{
Sasha Behrouzi\IEEEauthorrefmark{1},
Lichao Wu\IEEEauthorrefmark{2},
Mohamadreza Rostami\IEEEauthorrefmark{1}, and
Ahmad-Reza Sadeghi\IEEEauthorrefmark{1}
}
\IEEEauthorblockA{\IEEEauthorrefmark{1}Technical University of Darmstadt\\
\{sasha.behrouzi, mohamadreza.rostami\}@trust.tu-darmstadt.de\\
ahmad.sadeghi@trust.informatik.tu-darmstadt.de
}
\IEEEauthorblockA{\IEEEauthorrefmark{2}University of Bristol\\
lichao.wu@bristol.ac.uk}
}

\maketitle

\input{Sections_submission/0_abstract}

\IEEEpeerreviewmaketitle

\input{Sections_submission/1_introduction}
\input{Sections_submission/2_background}

\input{Sections_submission/3_design}

\input{Sections_submission/4_implementation}

\input{Sections_submission/5_results}

\input{Sections_submission/6_ablation}

\input{Sections_submission/7_related_works}

\input{Sections_submission/8_conclusion}

\bibliographystyle{IEEEtran}
\bibliography{bibliography}

\input{Sections_submission/10_appendix}

\end{document}

%% file: macros.tex
\usepackage{pifont}
\usepackage[dvipsnames]{xcolor}
	
\usepackage{color, colortbl}

\newcommand{\ournameNoSpace}{NeST} 
\newcommand{\ourname}{\ournameNoSpace\xspace}

\newcommand*\colourcheck[1]{%
  \expandafter\newcommand\csname #1check\endcsname{\textcolor{#1}{\ding{52}}}%
}

\newcommand*\colourcross[1]{%
  \expandafter\newcommand\csname #1cross\endcsname{\textcolor{#1}{\ding{55}}}%
}

\definecolor{LightCyan}{rgb}{0.88,1,1}
	
\definecolor{Gray}{gray}{0.8}
\colourcheck{green}
\colourcross{red}

\definecolor{darkgreen}{rgb}{0.0, 0.5, 0.0}

%% file: Sections_submission/0_abstract.tex
\begin{abstract}

Safety alignment is essential for the responsible deployment of Large Language Models (LLMs). Yet, existing approaches often rely on heavyweight fine-tuning that is costly to update, audit, and maintain across model families. Full fine-tuning incurs substantial computational and storage overhead, while parameter-efficient methods, e.g., Low-Rank Adaptation (LoRA), trade efficiency for inconsistent safety gains and sensitivity to design choices. Safety intervention mechanisms reduce unsafe outputs without modifying model weights, but do not directly shape or preserve the internal representations that govern safety behavior. These limitations hinder rapid and reliable safety updates, particularly in settings where models evolve frequently or must adapt to new policies and domains.

We present NeST, a Neuron-Selective Tuning framework for efficient post-hoc safety alignment. NeST identifies safety-relevant feed-forward neurons via activation probing on vanilla harmful and benign prompts, clusters neurons with similar activation profiles, and trains shared cluster-level updates while freezing the rest of the model. Importantly, NeST is trained only on vanilla malicious prompts, without using jailbreak-specific attack data, yet generalizes robustly to diverse jailbreaks. The learned updates are then folded into the original weights, incurring no inference-time overhead. Evaluated on 14 open-weight language and multimodal models, NeST outperforms lightweight baselines and approaches full fine-tuning robustness with significantly fewer trainable parameters. On text-only models, NeST reduces average jailbreak attack success rate from 44.5\% to 1.1\% while training only 0.4M parameters on average. Across multimodal settings, it reduces ASR from 55.3\% to 1.1\%, and for downstream fine-tuned variants, it restores safety by reducing ASR from 53.8\% to 0.8\%. These results show that robust, maintainable safety alignment can be achieved by concentrating adaptation on localized, functionally coherent safety structures.

\end{abstract}

%% file: Sections_submission/1_introduction.tex
\section{Introduction}
\label{sec:introduction}

Large Language Models (LLMs) have become a core component of modern AI systems, supporting text generation, translation, question answering, reasoning, and conversational interaction~\cite{matarazzo2025survey,minaee2024large,kaddour2023challenges}. Despite these capabilities, deployed LLMs remain vulnerable to jailbreak attacks that elicit harmful or policy-violating outputs through carefully crafted inputs~\cite{wei2023jailbroken,shen2024anything,yu2024llm,chang2024play}. As LLMs are increasingly deployed via hosted APIs and adapted for downstream applications, LLM service providers need safety alignment methods that are \emph{effective}, \emph{lightweight}, and \emph{maintainable} as models, safety policies, and attack strategies evolve.

\noindent\textbf{Safety Alignment and Its Limitations.} Existing safety-alignment methods satisfy these requirements only partially. Output-level alignment methods~\cite{ouyang2022training,rafailov2023direct}, including widely adopted approaches such as reinforcement learning from human feedback (RLHF)~\cite{ouyang2022training} and preference-based optimization~\cite{rafailov2023direct}, improve safety by optimizing models to produce preferred, often refusal-style, responses. While effective at reducing unsafe generations, these methods typically treat the parameter space as largely homogeneous and update broad regions of the model. As a result, safety alignment remains expensive and, even worse, unreliable when models are adapted to new downstream tasks~\cite{qi2023fine}. Repeating safety alignment is computationally costly and imposes a ``safety tax''~\cite{huang2025safety}, where stronger safety alignment comes at the expense of general model capability.
Parameter-efficient fine-tuning (PEFT) methods reduce this computational burden, but they do not resolve the underlying structural mismatch. Low-Rank Adaptation (LoRA)~\cite{hu2022lora}, for example, freezes the pretrained weights and injects trainable low-rank matrices into Transformer layers. However, it still operates over generic layers or projections rather than safety-relevant circuits, neurons, or representations. Consequently, it remains largely agnostic to the internal mechanisms that implement refusal behavior and harmfulness recognition. This limitation manifests in both directions: LoRA-based safety alignment can still harm general capabilities~\cite{zhou2025lssf}, while general LoRA-based fine-tuning can weaken or disrupt existing safety alignment, even when the downstream data is benign~\cite{hsu2024safe}. 
To reliably identify, preserve, or control the mechanisms responsible for safety,
recent work has begun to shift attention toward safety-relevant internal structures, but existing approaches remain limited. Structure-aware methods can be costly and still provide limited robustness against jailbreak attacks~\cite{zhao2025understanding}. Representation-level defenses take a different approach by intervening directly on internal activations to suppress harmful generation trajectories~\cite{zou2406improving}. However, because these methods are typically applied as inference-time controls, they introduce additional runtime overhead and complicate deployment. Taken together, existing methods are often expensive, structurally blind, or intervention-dependent, pointing to a missing capability in current safety alignment.

\noindent\textbf{Requirements for Practical Safety Alignment.} We argue that a practical post-hoc safety alignment method should satisfy three requirements: (i) \emph{parameter-efficient}: it should make safety updates cheap to train; (ii) \emph {structure-aware}: it should align updates with the model’s safety-relevant structure, rather than modifying broad parameter regions unrelated to refusal behavior; and (iii) \emph{maintainable}: it should produce a compact safety prior that can be reused to harden downstream fine-tuned variants. Recent studies support this direction by showing that safety-related behavior is not uniformly distributed across all parameters, but is often associated with localized internal components such as neurons, layers, and activation subspaces~\cite{mu2020compositional,antverg2021pitfalls,wu2025neurostrike,wu2025gatebreaker}. This suggests that safety alignment need not modify the entire model. Instead, adaptation capacity can be concentrated on the internal components most relevant to refusal behavior while leaving the rest of the model largely unchanged. This insight motivates our work.

\noindent\textbf{Our Goal and Contribution.} In this work, we introduce \ourname{}, an \emph{Ne}uron-\emph{S}elective \emph{T}uning framework for efficient post-hoc safety alignment. \ourname{} is designed for hosted deployment settings, where the model provider has offline access to model internals for safety tuning, while external attackers interact with the deployed model only through prompts. Rather than updating the full model or adding inference-time controls, \ourname{} identifies safety-relevant neurons from internal activations, clusters functionally similar safety neurons, and learns a compact shared update for each cluster while freezing all other parameters. After training, the learned updates are merged into the model weights, yielding a standard model with no additional inference-time cost. Since these updates are localized to a small set of safety-relevant neurons, they can be extracted as a compact safety prior and reused to harden downstream fine-tuned models, enabling efficient safety restoration through lightweight recovery training.
Across 14 state-of-the-art LLMs with different input modalities, \ourname{} substantially reduces attack success rates against jailbreaks while preserving core model capabilities. It achieves this with orders-of-magnitude fewer trainable parameters than full fine-tuning and state-of-the-art baselines. These results show that safety alignment can be made parameter-efficient, structure-aware, and maintainable by concentrating updates on localized internal components associated with refusal behavior. Our key contributions are summarized as follows:
\begin{itemize}
    \item We introduce \ourname{}, a \emph{Ne}uron-\emph{S}elective \emph{T}uning framework for post-hoc safety alignment. \ourname{} aligns parameter updates with localized safety-relevant components, enabling efficient safety adaptation without model-wide fine-tuning or inference-time intervention.
    \item We propose a cluster-based safety adaptation mechanism that groups safety-relevant neurons by activation similarity and learns shared updates within each group. This design provides coherent, structured modification of refusal behavior while freezing the remainder of the model.
    \item We show that \ourname{} achieves strong safety robustness across 14 open-weight LLMs spanning multiple model families and sizes. \ourname{} reduces the average attack success rate from 44.5\% to 1.1\%, achieving performance comparable to full fine-tuning and outperforming all other counterparts under our unified evaluation protocol. \ourname{} is also highly parameter-efficient, reducing the number of trainable parameters by more than 5,800$\times$ compared with full fine-tuning and by more than 3$\times$ compared with the lowest-rank LoRA baseline, while preserving core reasoning and knowledge capabilities.
    \item We evaluate \ourname{} across diverse inference settings and input modalities, including text-only, image-only, and reasoning-augmented multimodal generation. Across four multimodal LLMs, \ourname{} reduces average attack success rate from $55.3\%$ to $1.1\%$.
    \item We demonstrate that \ourname{} supports post-hoc downstream safety hardening by reusing safety-relevant neuron clusters identified in the base model. Across ten downstream fine-tuned variants, \ourname{} reduces average ASR from 53.8\% to 0.8\%, enabling efficient safety restoration after task-specific adaptation.
\end{itemize}

The remainder of this paper is structured as follows. Section~\ref{sec:preliminaries} provides background. Section~\ref{sec:framework} describes the design of \ourname{}, and Section~\ref{sec:Implementation} presents implementation details. Section~\ref{sec:case study} analyzes the structure of safety neurons. Section~\ref{sec:Performance Evaluation} evaluates \ourname{} against state-of-the-art baselines. Section~\ref{sec:ablation study} studies key hyperparameters. Section~\ref{sec:discussion} discusses broader implications and limitations. Section~\ref{sec:related} reviews related work, and Section~\ref{sec:conclusions} concludes this paper.

The artifact is available at the archival repository, \url{https://anonymous.4open.science/r/nestSP27-1C20}.

%% file: Sections_submission/2_background.tex
\section{Preliminaries}
\label{sec:preliminaries}

\subsection{Large Language Models}

Large language models are commonly implemented as decoder-only transformer architectures trained to model the conditional distribution of text. Given an input sequence, token representations are refined through a stack of transformer blocks. Each block applies self-attention followed by a position-wise feed-forward network (FFN), with residual connections around both components. Let $h_\ell \in \mathbb{R}^{d}$ denote the hidden representation at layer $\ell$. Abstracting away layer normalization for clarity, one transformer block updates the representation as:
\begin{equation}
\tilde{h}_\ell = h_\ell + \mathrm{Attn}(h_\ell), 
\qquad
h_{\ell+1} = \tilde{h}_\ell + \mathrm{FFN}(\tilde{h}_\ell),
\label{eq:transformer_block}
\end{equation}
where $\mathrm{Attn}(\cdot)$ denotes self-attention and $\mathrm{FFN}(\cdot)$ is applied independently to each token position~\cite{vaswani2017attention}.

Modern LLMs typically use gated feed-forward networks. For a token representation $x \in \mathbb{R}^{d}$, written as a row vector, the FFN transformation is:
\begin{equation}
\mathrm{FFN}(x) =
\big[\sigma(x W_{\mathrm{gate}}) \odot (x W_{\mathrm{up}})\big] W_{\mathrm{down}},
\label{eq:gated_fnn}
\end{equation}
where $W_{\mathrm{gate}}, W_{\mathrm{up}} \in \mathbb{R}^{d \times d_{\mathrm{ff}}}$ project the input into a higher-dimensional feed-forward space, $W_{\mathrm{down}} \in \mathbb{R}^{d_{\mathrm{ff}} \times d}$ maps the result back to the model dimension, $\sigma(\cdot)$ is a nonlinear activation such as SiLU, and $\odot$ denotes element-wise multiplication. This gated structure enables token-dependent activation of FFN dimensions, which we refer to as neurons. In \ourname{}, these FFN neurons are the units on which safety-neuron detection, clustering, and selective tuning operate. Such implementations are widely used in, e.g., GPT~\cite{achiam2023gpt}, LLaMA~\cite{touvron2023llama}, Qwen~\cite{yang2025qwen3}, and Gemma~\cite{team2025gemma}.

\subsection{LLM Fine-Tuning}

Fine-tuning adapts a pretrained LLM to a target task by optimizing its parameters on task-specific data. Let $\theta_0$ denote the pretrained parameters and $\mathcal{D}$ denote a dataset of input--output pairs $(x,y)$. LLM fine-tuning solves:
\begin{equation}
\theta^\star = \arg\min_{\theta} \; 
\mathbb{E}_{(x,y)\sim \mathcal{D}}
\left[ \mathcal{L}(f(x;\theta), y) \right],
\label{eq:ft-equation}
\end{equation}
where $f(x;\theta)$ is the model output and $\mathcal{L}$ is the training loss. In safety alignment, $\mathcal{D}$ typically contains harmful and benign prompts paired with desired safe responses: refusals for harmful requests and helpful answers for benign requests. 

Full fine-tuning updates all model parameters, which is computationally expensive for large models.
Parameter-efficient fine-tuning (PEFT) reduces this cost by freezing the pretrained weights and training only a small set of additional or restricted parameters. LoRA~\cite{hu2022lora}, for instance, augments a frozen linear transformation $W \in \mathbb{R}^{d_{\mathrm{in}} \times d_{\mathrm{out}}}$ with a trainable low-rank update:
\begin{equation}
x_{\mathrm{out}} = x_{\mathrm{in}}\left(W + \gamma AB\right),
\label{eq:lora}
\end{equation}
where $x_{\mathrm{in}} \in \mathbb{R}^{d_{\mathrm{in}}}$ is a row vector, $A \in \mathbb{R}^{d_{\mathrm{in}} \times r}$, $B \in \mathbb{R}^{r \times d_{\mathrm{out}}}$, $r \ll \min(d_{\mathrm{in}}, d_{\mathrm{out}})$, and $\gamma$ is a scaling factor. Only $A$ and $B$ are optimized, enabling adaptation with far fewer trainable parameters than full fine-tuning. \ourname{} follows the same broad goal of reducing trainable parameters, but differs in how the update space is chosen: instead of applying low-rank updates at the layer or projection level, it restricts updates to safety-relevant structures.

%% file: Sections_submission/3_design.tex
\section{\ourname}
\label{sec:framework}

\subsection{Threat Model}
\label{subsec:threat model}

\noindent\textbf{Adversary.}
We target a hosted black-box deployment setting in which attackers interact with the deployed model through prompts. This setting reflects common API-based LLM deployments, where the model provider controls the model internals while external users can issue queries and observe outputs. The adversary's objective is to elicit harmful, unsafe, or policy-violating responses by crafting adversarial inputs, including jailbreak prompts, indirect requests, and role-playing scenarios. The adversary may know that \ourname{} is deployed and may understand its high-level design, but does not have access to model parameters, gradients, activations, or the training and fine-tuning process. Stronger adversaries with white-box or training-time privileges are outside this deployment setting and are discussed in Section~\ref{sec:discussion}.

\noindent\textbf{Defender.}
The defender is the model provider or deployer who has access to the model architecture and parameters during offline training or fine-tuning. The defender's goal is to improve robustness against prompt-based safety violations while preserving general-purpose capabilities. To this end, the defender may analyze internal representations offline and apply structured safety-oriented training interventions before deployment. \ourname{} operates entirely at training time and introduces no inference-time interventions.

\subsection{Idea and High-Level Design}
\label{subsec:design}
\ourname{} is based on a simple intuition: safety behavior is not implemented uniformly across all model parameters. Instead, refusal behavior is often associated with localized internal components. \ourname{} leverages this structure to turn safety hardening into a targeted update problem: identify the neurons most associated with safety behavior, group them into functionally coherent clusters, and train shared updates only for those clusters while freezing the rest of the model.

\begin{figure*}[t]
\centerline{\includegraphics[width=\linewidth]{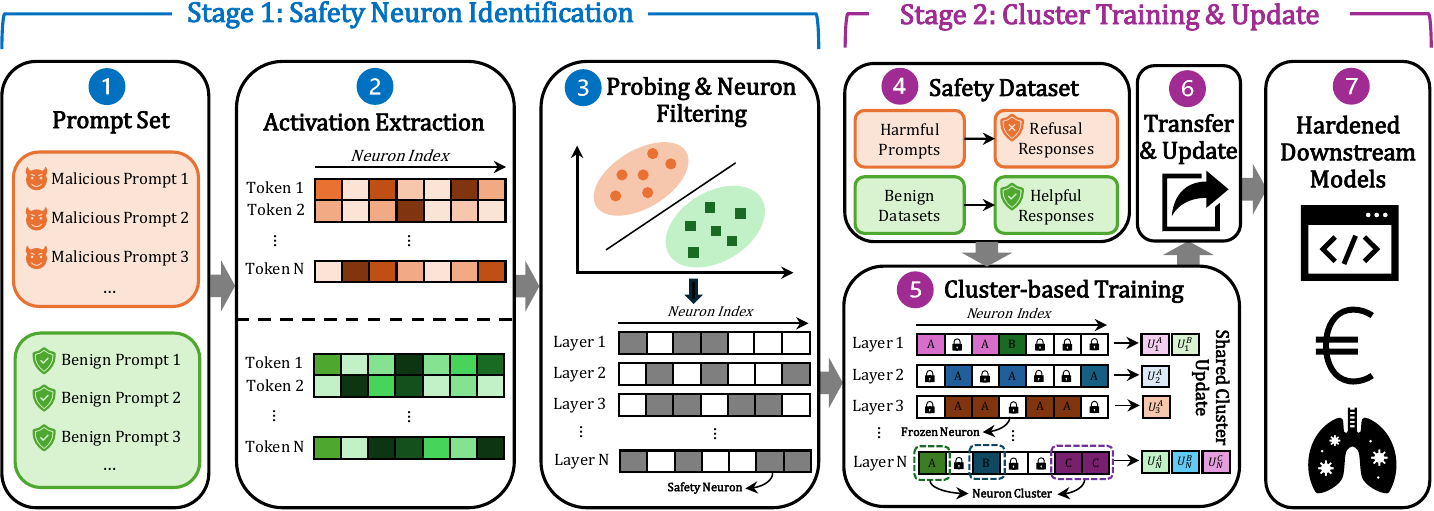}}
\caption{Overview of \ourname{}. Stage 1 identifies safety-relevant neuron clusters in the target model using activation differences between harmful and benign prompts, followed by probing and neuron filtering. Stage 2 trains shared cluster-level safety updates while freezing all non-selected neurons. These base-model clusters are transferable: the same selected neurons and cluster assignments are used to harden downstream models.}
\label{fig:overview_black}
\end{figure*}

Figure~\ref{fig:overview_black} shows the two-stage workflow of \ourname{}. In the first stage, \ourname{} identifies safety neurons. Given a prompt set containing both malicious (e.g., ``How to make a bomb?'') and benign (e.g., ``What is LLM?'') prompts, we collect neuron-level activations from the feed-forward modules of a safety-aligned model. A lightweight probing and filtering step then selects neurons whose activations are most discriminative of safety-relevant behavior. The output of this stage is a sparse set of safety neurons across layers.

In the second stage, \ourname{} performs cluster-based safety training. Using a safety dataset that pairs malicious prompts with refusal responses (e.g., ``Sorry, I can't...'') and benign prompts with helpful responses (e.g., ``Sure, here is...''), \ourname{} trains only the selected safety neurons. Safety neurons with similar activation behavior are assigned to the same cluster and share a trainable cluster update. All other neurons remain frozen. This design concentrates learning capacity on safety-relevant components while enforcing coherent updates among neurons that play similar roles. For downstream models, \ourname{} reuses the safety prior, i.e., selected safety neurons and cluster assignments, identified in the base model, then performs lightweight recovery training only on the corresponding cluster-level updates. This allows safety to be restored post hoc without rediscovering safety neurons or updating broad regions of the downstream model.

\subsection{Safety Neuron Identification}
\label{subsec:Safety Neurons Detection}

The first step of \ourname{} is to localize the internal components that participate in safety behavior. Specifically, we identify neurons in the frozen model whose activations already encode information relevant to the harmful-versus-benign distinction. 
Let $f_\theta$ denote a safety-aligned language model, and let $h_{\ell,t}(x) \in \mathbb{R}^d$ be the hidden representation at layer $\ell$ and token position $t$ for an input prompt $x$. A prompt may trigger safety behavior only at a small number of salient token positions, such as tokens describing the harmful intent. Using an average over all tokens could dilute such localized signals. We therefore summarize each neuron by its strongest response over the prompt using coordinate-wise max pooling:
\begin{equation}
\tilde{h}_{\ell,i}(x) = \max_{t \in \mathrm{tokens}(x)} h_{\ell,t,i}(x),
\quad i \in \{1,\ldots,d\},
\label{eq:pooled_activation}
\end{equation}
where $\tilde{h}_{\ell}(x) \in \mathbb{R}^d$ is the prompt-level activation vector.

Next, we use a linear probe to distinguish malicious prompts from benign prompts based on their neuron activations. The probe is intentionally simple: a linear model identifies neurons whose activations are directly predictive of safety-critical inputs, while avoiding the use of a high-capacity detector that could obscure which individual neurons carry the signal. Let malicious prompts be labeled as the positive class. For each layer $\ell$, the probe predicts
\begin{equation}
p_\ell(x) = \sigma\!\left(w_\ell^\top \tilde{h}_\ell(x) + b_\ell\right),
\label{eq:linear_probe}
\end{equation}
where $w_\ell \in \mathbb{R}^d$ and $b_\ell \in \mathbb{R}$ are the probe parameters, and $\sigma(\cdot)$ is the sigmoid function. The learned coefficient $w_{\ell,i}$ provides a neuron-level safety score: a larger positive value indicates that higher activation of neuron $i$ increases the probe's confidence that the prompt is harmful and therefore safety-critical.
To select a sparse and comparable set of safety neurons across layers, we standardize probe weights within each layer:
\begin{equation}
z_{\ell,i} =
\frac{w_{\ell,i} - \mu_\ell}{\sigma_\ell + \epsilon},
\label{eq:z_score}
\end{equation}
where $\mu_\ell$ and $\sigma_\ell$ are the mean and standard deviation of $\{w_{\ell,i}\}_{i=1}^d$, and $\epsilon$ is a small constant for numerical stability. This normalization accounts for layer-wise scale differences in probe weights. We then define the safety-neuron set as:
\begin{equation}
\mathcal{S}_\ell =
\left\{ i \;\middle|\; z_{\ell,i} > z_{\mathrm{thr}} \;\wedge\; w_{\ell,i} > 0 \right\},
\label{eq:safety_neurons}
\end{equation}
where $z_{\mathrm{thr}}$ controls the selectivity of detection. The threshold keeps only neurons whose association with safety-critical prompts is unusually strong relative to other neurons in the same layer, while the positivity constraint ensures that selected neurons are aligned with the harmful/refusal-triggering class rather than benign behavior.

\subsection{Neuron Clustering}
\label{subsec:Clustering Safety Neurons}

The safety neurons detected in Section~\ref{subsec:Safety Neurons Detection} need not play identical roles. Some may respond broadly to harmful intent, while others may specialize in particular prompt patterns or refusal-triggering contexts. Updating all safety neurons independently would increase the number of trainable parameters and could lead to unstable, uncoordinated changes. Conversely, forcing all safety neurons in a layer to share a single update may underfit heterogeneous safety behavior. \ourname{} uses clustering as a middle ground: neurons with similar activation profiles across prompts share an update, while neurons with different response patterns can be adapted separately. Formally, let $\mathcal{S}_\ell = \{ i_1, \dots, i_m \}$ denote the indices of the $m$ safety neurons selected at layer $\ell$. For a prompt set $\mathcal{X}=\{x_1,\dots,x_N\}$ containing both harmful and benign prompts, let $\tilde{h}_\ell(x) \in \mathbb{R}^d$ be the max-pooled activation vector defined in Eq.~\eqref{eq:pooled_activation}. We represent each safety neuron $i_j$ by its response profile across the prompt set:
\begin{equation}
\mathbf{a}_{\ell,i_j}
=
\left[
\tilde{h}_{\ell,i_j}(x_1), \dots, \tilde{h}_{\ell,i_j}(x_N)
\right]^\top
\in \mathbb{R}^N .
\label{eq:activation_profile}
\end{equation}
This vector captures how neuron $i_j$ responds across diverse safety-relevant and benign contexts. Two neurons with similar profiles are activated by similar prompts and are therefore likely to support related safety behavior.
Next, we cluster
$\{\mathbf{a}_{\ell,i_j}\}_{j=1}^m$
using $k$-means. For a candidate number of clusters $k$, we solve:
\begin{equation}
\min_{\{C_1,\dots,C_k\}}
\sum_{c=1}^k
\sum_{i_j \in C_c}
\left\|
\mathbf{a}_{\ell,i_j} - \boldsymbol{\mu}_c
\right\|_2^2,
\label{eq:kmeans}
\end{equation}
where $C_c$ is the set of neurons assigned to cluster $c$ and $\boldsymbol{\mu}_c$ is the corresponding centroid. 

The number of clusters controls the trade-off between expressiveness and parameter efficiency. A larger $k$ allows more specialized updates but introduces more trainable parameters, while a smaller $k$ enforces stronger sharing. We select $k$ based on the activation geometry. Specifically, we evaluate $k \in \{1,2,\dots,k_{\max}\}$, where $k_{\max}$ equals the number of neurons to be clustered. For $k\ge 2$, we compute the silhouette score:
\begin{equation}
\mathrm{Silhouette}(k)
=
\frac{1}{m}
\sum_{j=1}^m
\frac{b(i_j)-a(i_j)}
{\max\{a(i_j),b(i_j)\}},
\label{eq:silhouette}
\end{equation}
where $a(i_j)$ is the mean distance from neuron $i_j$ to other neurons in its assigned cluster, and $b(i_j)$ is the minimum mean distance from neuron $i_j$ to neurons in any other cluster. 
We choose the $k\ge 2$ with the highest silhouette score only if the score exceeds a threshold $\gamma$; otherwise, we use $k=1$ and treat the selected safety neurons in layer $\ell$ as a single group.
The final clustering defines a mapping:
\begin{equation}
c_\ell : \mathcal{S}_\ell \rightarrow \{0,\dots,k_\ell-1\},
\label{eq:cluster_mapping}
\end{equation}
which assigns each safety neuron to a cluster. This mapping determines parameter sharing during LLM tuning: neurons in the same cluster receive the same trainable update, while neurons in different clusters can be adapted separately.

\subsection{Clustered Training \& Downstream Hardening}
\label{subsec:neuron selective tuning}

After identifying and clustering safety neurons, \ourname{} performs safety tuning by restricting parameter updates to those clustered neurons. The rationale is to modify only the components most associated with refusal behavior, while leaving the rest of the model unchanged to reduce unnecessary interference with general capabilities. 
Concretely, \ourname{} learns one shared update per cluster, so neurons with similar activation behavior are adapted coherently.
Consider a feed-forward projection matrix \(W_\ell \in \mathbb{R}^{d_{\mathrm{in}}\times d_{\mathrm{out}}}\) at layer \(\ell\), where each output dimension corresponds to one neuron. Let \(\mathcal{S}_\ell\) be the set of detected safety neurons and let \(c_\ell:\mathcal{S}_\ell \rightarrow \{0,\dots,k_\ell-1\}\) be the cluster assignment from Eq.~\eqref{eq:cluster_mapping}. For the \(k_\ell\) clusters, \ourname{} introduces a trainable cluster-update matrix $ U_\ell \in \mathbb{R}^{d_{\mathrm{in}} \times k_\ell}$, where each column of \(U_\ell\) is the shared update vector for one safety-neuron cluster. The base weight \(W_\ell\) remains frozen. During training, each safety neuron \(i \in \mathcal{S}_\ell\) uses the effective weight column:
\begin{equation}
W'_{\ell,:,i}
=
W_{\ell,:,i}
+
U_{\ell,:,c_\ell(i)},
\qquad i \in \mathcal{S}_\ell,
\label{eq:effective_weight_column}
\end{equation}
while non-safety neurons remain unchanged:
\begin{equation}
W'_{\ell,:,i}
=
W_{\ell,:,i},
\qquad i \notin \mathcal{S}_\ell.
\label{eq:frozen_weight_column}
\end{equation}

This reparameterization keeps the trainable parameter count proportional to \(d_{\mathrm{in}}k_\ell\), rather than \(d_{\mathrm{in}}d_{\mathrm{out}}\). Since \(k_\ell\) is small and only safety-neuron clusters receive updates, \ourname{} provides a low-dimensional safety adaptation space aligned with the model's internal safety structure. During supervised safety tuning, gradients are propagated only through the cluster-update parameters \(\{U_\ell\}\). After training, the learned updates are folded into the corresponding columns of \(W_\ell\). 

The same design also enables post-hoc hardening of downstream fine-tuned variants derived from the same base model. Since downstream variants preserve the base model architecture and neuron indexing, \ourname{} transfers the structural safety prior identified in the base model, namely the safety-neuron sets $\mathcal{S}$ and their cluster assignments $c$, to the downstream model. \ourname{} instantiates the same cluster-level update parameterization on the corresponding part of the downstream model, freezes all non-selected neurons, and trains only the downstream cluster-update parameters on safety data. The learned recovery updates are then folded into the downstream model weights. This restores safety post hoc while preserving the downstream task-specific weights outside the localized safety update space, yielding a safety-hardened downstream model.

%% file: Sections_submission/4_implementation.tex
\section{Implementation}
\label{sec:Implementation}

\subsection{Safety Neuron Detection}
\label{subsec:impl_safety_neuron_detection}

We construct a balanced and diverse probing dataset containing harmful and benign prompts. Harmful prompts are drawn from \texttt{CatHarmfulQA}~\cite{bhardwaj2024language}, \texttt{HarmfulQA}~\cite{bhardwaj2023redteaming}, and the \texttt{LLM-LAT harmful dataset}~\cite{harmful-dataset}. Benign prompts are sampled from the training split of \texttt{Natural Reasoning}~\cite{yuan2025naturalreasoningreasoningwild28m}, with the number of benign prompts matched to the total number of harmful prompts. We assign harmful prompts to the positive class and benign prompts to the negative class.
For each model, we register forward hooks on the feed-forward \texttt{gate\_proj} and \texttt{up\_proj} modules in every transformer block. We focus on FFN projections because they provide neuron-level dimensions that can be selectively updated and have been shown to encode localized behavioral features. For each prompt, we perform a single forward pass through the frozen model and record token-level activations from the hooked modules. Next, max-pooling is applied over the sequence dimension, as in Eq.~\eqref{eq:pooled_activation}, to produce a single prompt-level activation vector per layer and projection. The collected activations for all prompts form a matrix in \(\mathbb{R}^{N \times d}\), where \(N\) is the number of prompts and \(d\) is the FFN output dimension.
This matrix is used to train a linear probe to distinguish harmful from benign prompts using binary cross-entropy loss. 
The probe weights are standardized and, following Eq.~\eqref{eq:safety_neurons}, output a sparse set of safety neurons for each layer. The influence of the selection threshold $z_{thr}$ is studied in Section~\ref{subec:Impact of the z-Threshold}.
The stability of safety-neuron detection across different probe seeds and harmful datasets is evaluated in Section~\ref{subsec:Stability of Safety-Neuron Detection}.

\subsection{Safety Neuron Clustering}
\label{subsec:impl_safety_neuron_clustering}

For each layer, we only cluster the neurons selected during safety-neuron detection. We reuse the pooled activation matrix from the detection stage and extract the columns corresponding to the selected safety neurons, obtaining a matrix in \(\mathbb{R}^{N \times m}\), where \(m\) is the number of selected safety neurons. Each selected neuron is represented by one column of this matrix, i.e., its activation profile across the probing prompts.
We then apply \(k\)-means to the activation profiles, treating each safety neuron as one data point. For each layer-projection pair, we evaluate candidate clustering up to \(k_{\max}\) and select the clustering with the highest silhouette score, preferring smaller \(k\) in the case of ties. As discussed in Section~\ref{subsec:Clustering Safety Neurons}, if no candidate split exceeds the silhouette threshold \(\gamma\), we assign all selected safety neurons in that layer-projection pair to a single cluster. The resulting cluster mapping determines which neurons share update parameters and is fixed before (downstream) fine-tuning. In Section~\ref{subsec:Impact of Clustering Strength}, we study the impact of the neuron clustering.

\subsection{Neuron-Selective Tuning}
\label{subsec:impl_neuron_selective_tuning}

Before fine-tuning, all original model parameters are frozen. For each FFN projection containing selected safety neurons, \ourname{} introduces trainable cluster-level update parameters \(U_\ell \in \mathbb{R}^{d_{\mathrm{in}} \times k_\ell}\), where \(k_\ell\) is the number of clusters for that layer-projection pair. During the forward pass, each selected safety neuron receives the update associated with its cluster, while all non-selected neurons remain unchanged.
\ourname{} uses supervised safety data containing both harmful and benign examples. Harmful prompts are paired with refusal responses, while benign prompts are paired with helpful responses. This balanced objective strengthens refusal behavior without encouraging blanket refusal of benign requests. Training optimizes the standard language-modeling loss under the neuron-selective parameterization. After fine-tuning, the learned cluster updates are folded into the corresponding FFN projection weights, and the reparameterization modules are removed. The resulting model has the same architecture, parameterization, and inference cost as the original model, while the safety prior, i.e., safety neurons and clusters, is stored separately for hardening downstream fine-tuned variants of the same base model.

%% file: Sections_submission/5_results.tex
\section{Case Study}
\label{sec:case study}

\begin{figure}[t]
\centering
\subfloat[FNN gate projection]{
    \includegraphics[width=0.5\linewidth]{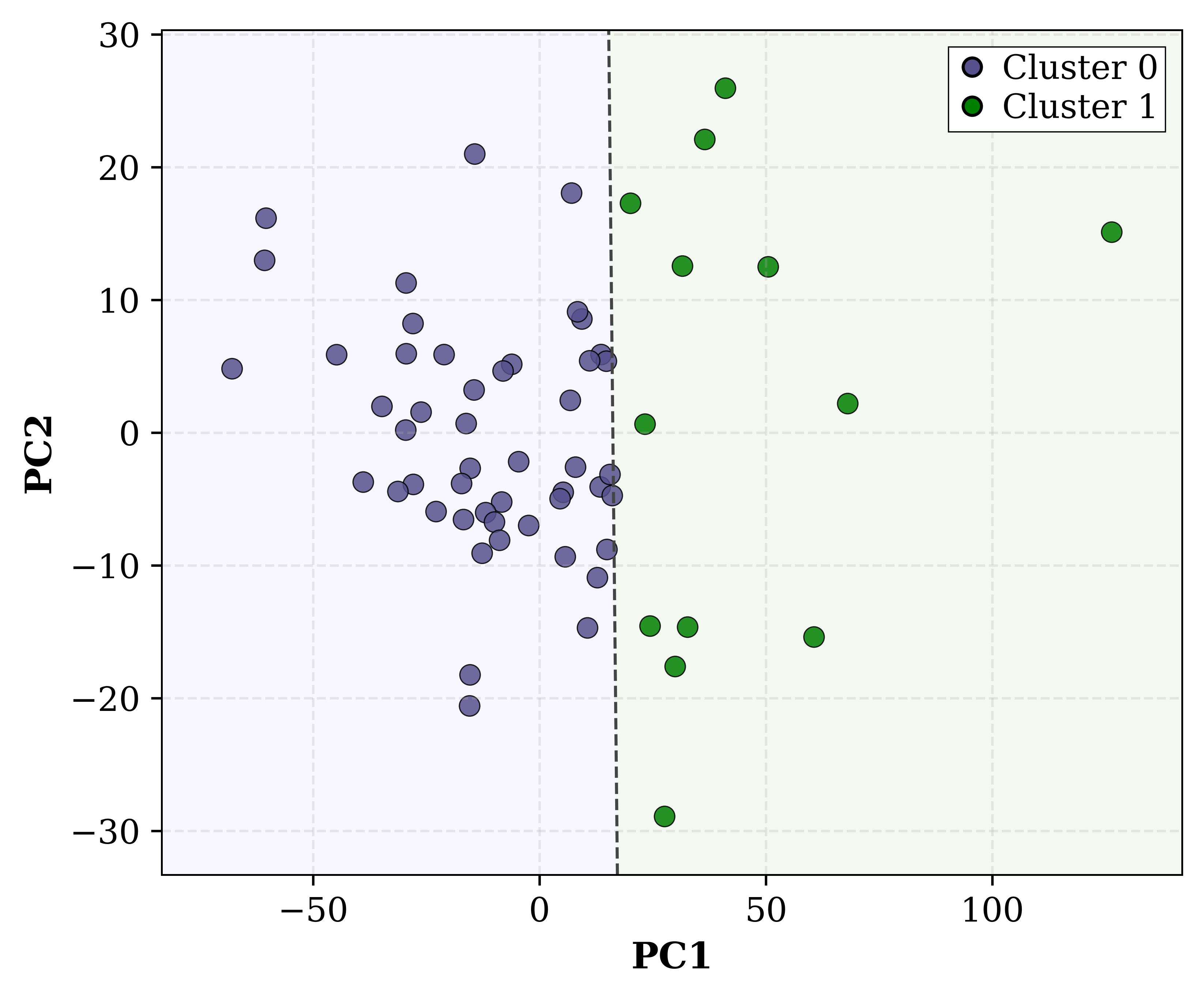}
    \label{fig:cluster_gate_l6}
}
\subfloat[FNN up projection]{
    \includegraphics[width=0.5\linewidth]{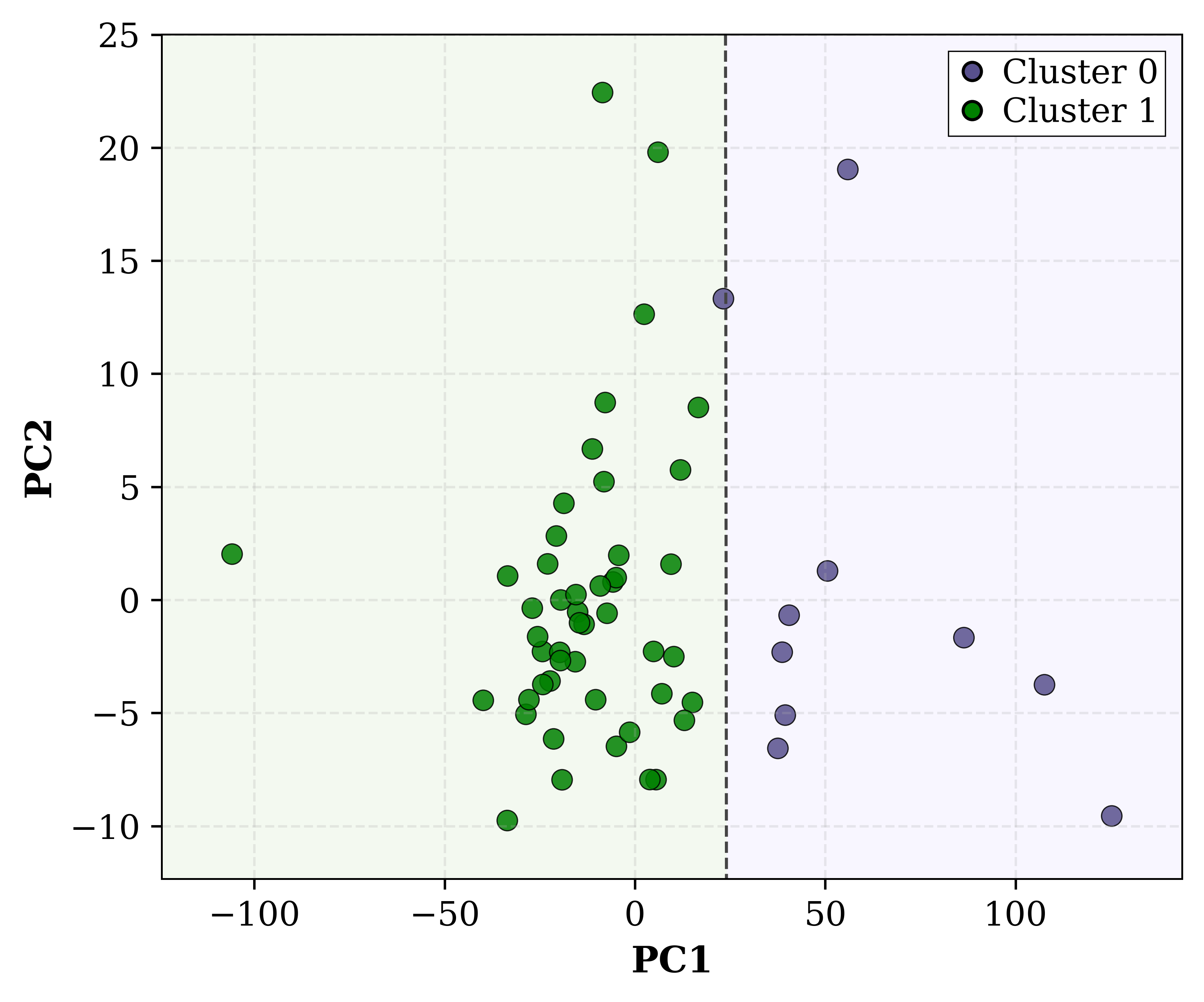}
    \label{fig:cluster_up_l6}
}
\caption{PCA visualization of clustered safety neurons in layer~6 of LLaMA-3.2-1B-Instruct.}
\label{fig:cluster_scatter_l6}
\end{figure}

This case study examines whether the safety structure used by \ourname{} is visible in model activations and relevant to optimization. We focus on LLaMA-3.2-1B-Instruct and analyze layer~6, a representative middle layer motivated by prior work showing that safety-alignment behavior often emerges in mid-network layers~\cite{li2024safety}. For this layer, we consider safety neurons detected in the FNN \texttt{gate\_proj} and \texttt{up\_proj} modules. For each detected neuron, we collect its activation profile across the probing prompt set, cluster neurons using the procedure in Section~\ref{subsec:Clustering Safety Neurons}, and project the resulting activation profiles into two dimensions using PCA. Each point in Figure~\ref{fig:cluster_scatter_l6} corresponds to a single safety neuron and is colored by its cluster assignment.

Figure~\ref{fig:cluster_scatter_l6} shows that safety neurons form compact groups in both projections. This indicates that the selected neurons are not merely a scattered set of independent units; rather, they exhibit structured activation patterns across prompts. This observation supports the cluster design of \ourname{}: neurons that respond similarly to safety-relevant inputs can share an update, allowing the method to reduce trainable parameters while preserving functional coherence.

\begin{figure}[t]
    \centering
    \includegraphics[width=\columnwidth]{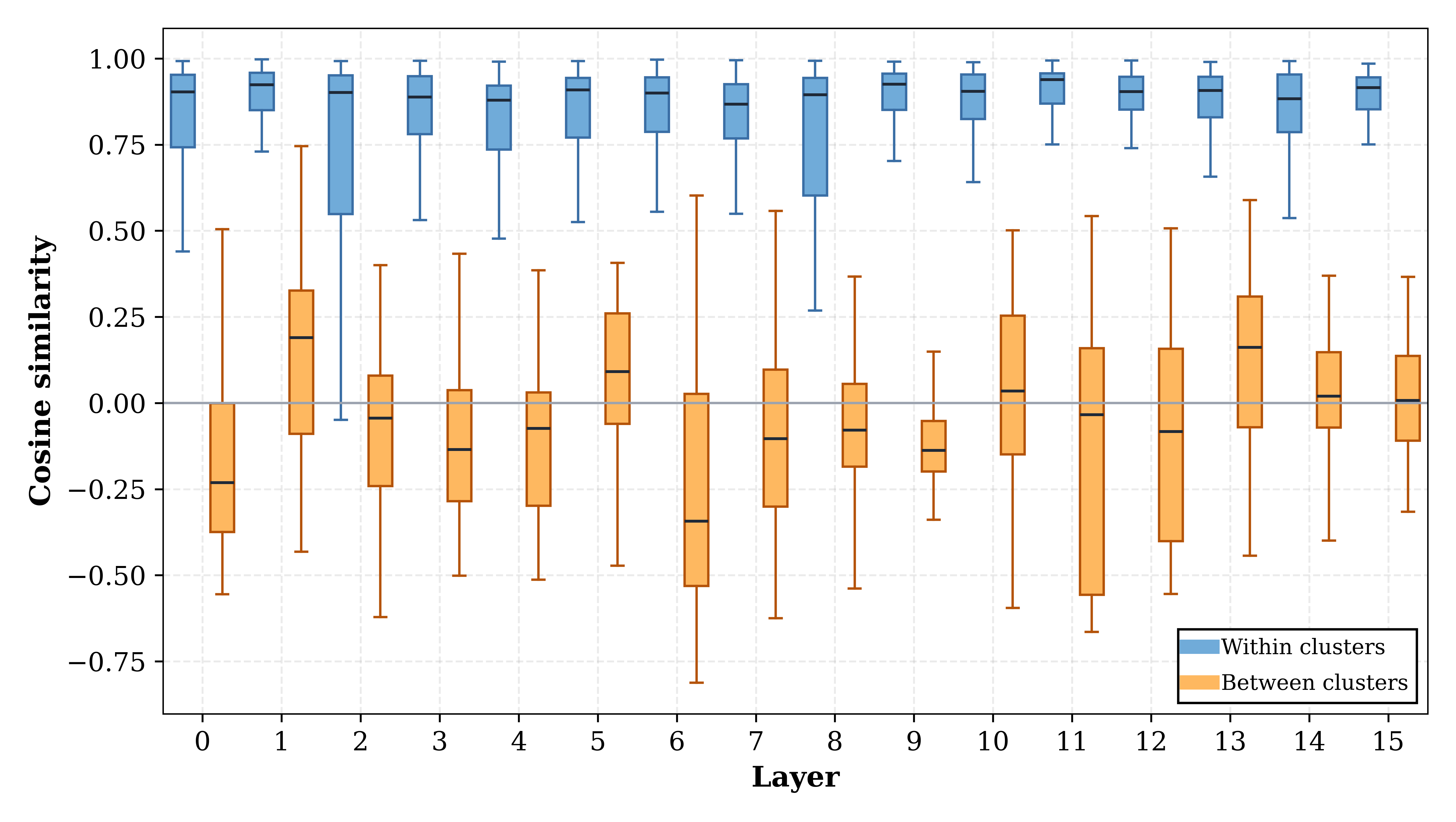}
    \caption{Gradient alignment of safety neurons across layers.}
    \label{fig:grad_align_layers}
\end{figure}
Next, we evaluate whether these clusters also reflect the optimization geometry encountered during fine-tuning. This is important because \ourname{} ties parameter updates within each cluster. If clustered neurons receive similar gradient directions during training, then shared cluster updates align with the model's natural update direction. During supervised safety tuning, we extract gradient vectors for the weights associated with detected safety neurons and compute pairwise cosine similarity between neuron gradients. We then compare pairs from the same cluster against pairs from different clusters.
As shown in Figure~\ref{fig:grad_align_layers}, within-cluster neuron pairs consistently have higher gradient cosine similarity than between-cluster pairs, which are substantially lower and often centered near zero. This separation indicates that neurons grouped by activation similarity also tend to receive aligned update directions during safety tuning. Consequently, the cluster-tied update rule in Eq.~\eqref{eq:effective_weight_column} matches the observed optimization structure: neurons that \ourname{} updates jointly are precisely those that would otherwise receive similar gradient updates. Together, the activation visualizations and gradient-alignment analysis provide empirical support for using clustered neuron-selective safety updates.

\section{Experimental Results}
\label{sec:Performance Evaluation}
\noindent\textbf{Models and Baselines.}
We evaluate \ourname{} on 14 open-weight instruction-tuned models spanning roughly 1B to 27B parameters and covering both text-only and multimodal settings. For text-only evaluation, we use models from four major families: Meta Llama, Alibaba Qwen2.5/Qwen3~\cite{qwen2.5,hui2024qwen2,yang2024qwen2,yang2025qwen3}, Microsoft Phi-4~\cite{abdin2024phi,abouelenin2025phi}, and Google Gemma~\cite{team2024gemma,team2025gemma}. For multimodal evaluation, we include vision-language variants from the Gemma and Qwen to test whether \ourname{} remains effective across input modalities and reasoning settings.
We compare \ourname{} against the original models and four safety adaptation baselines: full-parameter supervised fine-tuning (Full FT), LoRA~\cite{hu2022lora}, Circuit Breaker~\cite{zou2406improving}, and SN-Tune~\cite{zhao2025understanding}.

\noindent\textbf{Training Data.}
For safety training, we use only the vanilla malicious prompts (``How to make a bomb?'') from \texttt{WildJailbreak}~\cite{wildteaming2024}, excluding the corresponding jailbreak variants. These prompts express harmful intent directly, encouraging the model to generalize from direct harmful intent to more diverse jailbreak attacks.
Benign examples are sampled from the training split of \texttt{Natural Reasoning}~\cite{yuan2025naturalreasoningreasoningwild28m}, with the number of benign prompts matched to the harmful set. The resulting 20k-example dataset is shuffled and split into training and validation sets using a 95/5 split.
For Circuit Breaker, we follow the authors' original training setup, which uses a curated circuit-breaker dataset together with a retained set of benign and refusal-aligned examples. We report results under both the authors' original setting and our experimental setting. In our experimental setting, we augment the Circuit Breaker training data with additional \texttt{WildJailbreak} prompts while keeping the original retain set, so that this baseline is also exposed to the same safety-tuning distribution used by the other methods.

\noindent\textbf{Training Configurations.}
The original model is used as the baseline. Full FT updates all model parameters for five epochs. LoRA inserts rank-$1$ adapters into FNN projections and trains for five epochs; we use a small rank to keep the trainable parameter count close to \ourname{} and enable a direct parameter-efficiency comparison. Circuit Breaker follows original representation rerouting configurations, using rank-$32$ adapters, dropout $0.05$, bfloat16 precision, gradient checkpointing, and up to 150 optimization steps. \ourname{} freezes the base model and trains only the cluster-level updates for five epochs, under the same optimization settings as Full FT and LoRA.

\noindent\textbf{Evaluation Settings.}
We evaluate all methods on 1{,}000 held-out jailbreak prompts sampled from \texttt{WildJailbreak}. These evaluation prompts are adversarially phrased to bypass safety alignment through, for instance, role-playing, obfuscation, and indirect instruction. This creates a strict generalization test: methods must transfer from easy-to-reject malicious training prompts to harder jailbreak-style attacks that are more likely to evade standard refusal behavior. The evaluation set is disjoint from all data used for safety-neuron detection, clustering, and fine-tuning. To further evaluate whether \ourname{} generalizes across jailbreak benchmarks, Section~\ref{subsec:stability dataset_and_judge} reports additional results on \texttt{JailBreakV-28k}~\cite{luo2024jailbreakv28k}, a separate jailbreak dataset.

For each model and defense configuration, prompts are formatted using the model's native instruction template, and responses are generated with greedy decoding and model-specific maximum generation lengths. We measure attack success rate (ASR), defined as the fraction of prompts that produce unsafe, non-refusal responses. We use Qwen3Guard-Gen-8B~\cite{zhao2025qwen3guard} as the primary safety judge. A response is counted as a successful jailbreak if Qwen3Guard labels it unsafe and the output does not contain an explicit refusal. 
To verify that the results are not specific to a specific judge, we repeat the evaluation with Llama-Guard-3-8B~\cite{grattafiori2024llama} in Section~\ref{subsec:stability dataset_and_judge}. The same evaluation protocol is applied uniformly across all models and methods. 
Table~\ref{tab:setup} summarizes the training and evaluation setup.
\begin{table}[t]
\centering
\scriptsize
\setlength{\tabcolsep}{4pt}
\caption{
Training and evaluation configurations. Full FT, LoRA, and \ourname{} use the same balanced 20k training set. 
}
\begin{tabular}{lcccc}
\toprule
\textbf{Method} & \textbf{Scope} & \textbf{Train Data} & \textbf{Epochs/Steps} & \textbf{Eval. Data} \\
\midrule
Baseline & - & - & - & 1k jailbreak\\
Full FT & All params & 20k direct & 5 epochs & 1k jailbreak \\
LoRA & FNN & 20k direct & 5 epochs & 1k jailbreak \\
Circuit Breaker & FNN+ATTN & 25k & $\leq$150 steps & 1k jailbreak \\
\ourname{} & Neuron clusters & 20k direct & 5 epochs & 1k jailbreak \\
\bottomrule
\end{tabular}
\label{tab:setup}
\end{table}

\begin{table*}[t]
\centering
\caption{ASR and trainable parameters (Param., in millions) across target models and safety adaptation methods.}
\begin{tabular}{l|c|cc|ccc|cc|cc|cc}
\toprule
\multirow{2}{*}{\textbf{Target Model}}
& \textbf{Baseline}
& \multicolumn{2}{c|}{\textbf{Full FT}}
& \multicolumn{3}{c|}{\textbf{Circuit Breaker}}
& \multicolumn{2}{c|}{\textbf{SN-Tune}}
& \multicolumn{2}{c|}{\textbf{LoRA}}
& \multicolumn{2}{c}{\textbf{\ourname{}}} \\
\cmidrule(lr){2-2}
\cmidrule(lr){3-4}
\cmidrule(lr){5-7}
\cmidrule(lr){8-9}
\cmidrule(lr){10-11}
\cmidrule(lr){12-13}
& \textbf{ASR}
& \textbf{ASR} & \textbf{Para.}
& \textbf{ASR$_{\text{Orig.}}$} & \textbf{ASR$_{\text{Exp.}}$} & \textbf{Para.}
& \textbf{ASR} & \textbf{Para.}
& \textbf{ASR} & \textbf{Para.}
& \textbf{ASR} & \textbf{Para.} \\
\midrule
Llama-3.2-1B  & 40.5\% & 2.1\% & 1,235.8 & 35.8\% & 58.1\% & 8.5  & 35.9\% & 16.38 & 1.4\% & 3.3  & 0.3\% & 0.1 \\
Llama-3.2-3B  & 54.2\% & 1.9\% & 3,212.8 & 48.1\% & 74.8\% & 18.2 & 23.6\% & 43.00 & 3.4\% & 5.0  & 0.5\% & 0.3 \\
Qwen2.5-7B    & 69.4\% & 1.6\% & 7,615.6 & 38.2\% & 79.1\% & 30.3 & 67.6\% & 50.17 & 19.2\% & 0.5 & 0.6\% & 0.4 \\
Qwen2.5-14B   & 58.1\% & 0.8\% & 14,770.0 & 20.1\% & 65.2\% & 51.6 & 58.6\% & 122.8 & 13.0\% & 18.8 & 0.8\% & 1.0 \\
Phi-4-mini    & 25.7\% & 0.3\% & 3,836.0 & 24.6\% & 44.5\% & 13.4 & 34.2\% & 598.6 & 5.7\% & 1.6  & 0.7\% & 0.2 \\
Phi-4         & 25.2\% & 1.9\% & 14,660.0 & 23.6\% & 68.2\% & 68.4 & 56.1\% & 102.4 & 17.8\% & 0.6 & 1.3\% & 0.4 \\
gemma-2b-it   & 22.0\% & 0.1\% & 2,506.2 & 8.6\%  & 39.7\% & 14.2 & 37.3\% & 29.95 & 4.2\% & 6.6  & 0.2\% & 0.2 \\
gemma-7b-it   & 34.0\% & 0.4\% & 8,537.7 & 12.9\% & 40.6\% & 37.5 & 44.7\% & 43.00 & 6.6\% & 1.5  & 5.3\% & 0.4 \\
Qwen3-4B      & 65.1\% & 0.7\% & 4,022.5 & 32.0\% & 76.4\% & 19.3 & 82.2\% & 46.08 & 1.6\% & 0.9  & 0.0\% & 0.4 \\
Qwen3-14B     & 50.9\% & 0.6\% & 15,768.3 & 27.1\% & 72.1\% & 31.5 & 64.1\% & 102.40 & 2.8\% & 1.8  & 1.0\% & 0.9 \\
\midrule
\emph{Average}
& \emph{44.5\%}
& \emph{1.0\%} & \emph{7,616.5}
& \emph{27.1\%} & \emph{61.9\%} & \emph{29.3}
& \emph{50.4\%} & \emph{115.8}
& \emph{7.6\%} & \emph{4.1}
& \emph{1.1\%} & \emph{0.4} \\
\bottomrule
\end{tabular}

\label{tab:asr_param}
\end{table*}

We evaluate each method along two dimensions: safety effectiveness via ASR and parameter efficiency via the number of trainable parameters. As shown in Table~\ref{tab:asr_param}, the baseline (original) models remain vulnerable to jailbreak prompts, with an average baseline ASR of 44.5\%. 
Full fine-tuning achieves the lowest average ASR, reducing it to 1.0\%, but requires updating the entire model. Among more lightweight methods, LoRA lowers the average ASR to 7.6\%, but remains above 10\% on several larger models, including Qwen2.5-7B, Qwen2.5-14B, and Phi-4. Circuit Breaker achieves 27.1\% average ASR under the authors' original setup, but increases to 61.9\% under our experimental setting. This suggests that Circuit Breaker primarily suppresses harmful behaviors observed during training, rather than learning a more generalized safety representation. Consequently, its effectiveness degrades under distribution shifts and stronger jailbreak prompts that differ from the optimization setting used in the original method. 

SN-Tune achieves an average ASR of 50.4\%, performing worse than both LoRA and Circuit Breaker under our evaluation setting. In particular, SN-Tune exhibits high ASR on several Qwen and Phi models, including 82.2\% on Qwen3-4B and 67.6\% on Qwen2.5-7B, indicating limited robustness against stronger held-out jailbreak prompts. In comparison, \ourname{} reduces the average ASR from 44.5\% to 1.1\%, consistently improving over the baseline and outperforming LoRA and SN-Tune on most target models. Moreover, despite updating only 0.4M parameters on average, \ourname{} remains competitive with full fine-tuning and even achieves lower ASR on five of the ten models, namely Llama-3.2-1B, Llama-3.2-3B, Qwen2.5-7B, Phi-4, and Qwen3-4B.

The trainable parameters count in Table~\ref{tab:asr_param} highlights the cost of these improvements. Full FT trains 7.6B parameters on average, which makes repeated safety updates expensive as models, policies, or attack distributions evolve. Circuit Breaker, SN-Tune, and LoRA reduce this cost to 29.3M, 115.5M, and 4.1M trainable parameters, respectively, but their safety performance is less consistent. In particular, SN-Tune requires substantially more trainable parameters than other lightweight approaches while still exhibiting relatively high ASR across multiple models. \ourname{} requires only 0.4M trainable parameters on average, corresponding to approximately 17{,}700$\times$ fewer trainable parameters than Full FT, 269$\times$ fewer than SN-Tune, 68$\times$ fewer than Circuit Breaker, and 9.4$\times$ fewer than LoRA. Despite using a much smaller update space, \ourname{} achieves ASR comparable to full FT while attaining a lower ASR than all lightweight baselines.

\subsection{Different Modalities and Inference Settings}
In this section, we further evaluate the stability of \ourname under different inference settings and input modalities, including text-only, image-only (by hard-coding prompts to images), and their combinations with explicit reasoning. The other settings align with the previous section.

Table~\ref{tab:transfer moe neurons} reports the ASR of four multimodal models with and without \ourname.
Without \ourname, the base models exhibit consistently high ASR across all configurations, with an average ASR of 55.3\%. Strong multimodal models such as Gemma-3-27B and Qwen3-VL-8B show especially high safety risks in image-based settings, reaching up to 85.0\% ASR. This indicates that neither modality nor inference style (e.g., reasoning-enabled decoding) mitigates the attack.
\begin{table}[ht]
\centering
\footnotesize
\caption{ASR with different modalities and inference settings.}
\begin{tabular}{ll|cc}
\toprule
\textbf{Target Model} & \textbf{Infer. Setting} & \textbf{w/o \ourname}  & \textbf{w/ \ourname}\\
\midrule
\multirow{4}{*}{\centering gemma-3-12b} & Text & 52.6\%  & 4.4\% \\
& Text \& Reasoning & 52.4\%  & 5.1\% \\
& Image & 40.8\%  & 0.2\% \\
& Image \& Reasoning & 39.8\%  & 1.0\% \\
\midrule
\multirow{4}{*}{\centering gemma-3-27b} & Text & 57.1\%  & 0.2\% \\
& Text \& Reasoning & 58.5\%  & 0.2\% \\
& Image & 84.2\%  & 0.0\% \\
& Image \& Reasoning & 85.0\%  & 0.0\% \\
\midrule
Qwen3-VL-8B & Text & 43.9\%  & 0.0\% \\
(Instruct) & Image & 84.0\%  & 0.2\% \\
\midrule
Qwen3-VL-8B & Text \& Reasoning & 43.1\%  & 1.2\% \\
(Thinking) & Image \& Reasoning & 22.2\%  & 1.2\% \\
\midrule
\multicolumn{2}{c|}{\emph{Average}}  & \emph{55.3\%} & \emph{1.1\%} \\
\bottomrule
\end{tabular}
\label{tab:transfer moe neurons}
\end{table}

In contrast, enabling \ourname leads to a dramatic and consistent reduction in ASR across all models, modalities, and inference settings. The average ASR drops to only 1.1\%, representing a relative reduction of over 54.2\%. Notably, this suppression remains effective for both text and image inputs, as well as for reasoning-augmented inference, where attacks are often more difficult to control due to longer generation trajectories.
These results suggest that \ourname{} remains stable and robust across heterogeneous inference conditions. Since its effectiveness does not depend on a specific input modality or decoding strategy, \ourname{} is well suited for deployment in multimodal and reasoning-enabled LLM systems. In particular, the near-zero ASR observed on larger models such as Gemma-3-27B indicates that the defense continues to provide strong protection as model size and inference complexity increase.
\begin{table*}[ht]
\centering
\caption{ASR before and after downstream fine-tuning, with and without post-hoc \ourname{} hardening.}
\begin{tabular}{lll|ccc}
\toprule
\textbf{Base Model} & \textbf{Downstream (Target) Model} & \textbf{Downstream Task} & \textbf{Baseline} & \textbf{Downstream w/o \ourname{}} & \textbf{Downstream w/ \ourname{}}\\
\midrule
Llama-3.2-1B & Vikhr-Llama-3.2-1B-Instruct & Russian language & 40.5\% & $32.1\%_{\textcolor{Green}{-8.4\%}}$ & $0.9\%_{\textcolor{Green}{-39.6\%}}$ \\
Llama-3.2-3B & Llama-Doctor-3.2-3B-Instruct & Medical consultation & 54.2\% & $64.7\%_{\textcolor{red}{+10.5\%}}$ & $0.4\%_{\textcolor{Green}{-53.8\%}}$ \\
Qwen2.5-7B & Fin-R1 & Financial reasoning & 69.4\% & $72.2\%_{\textcolor{red}{+2.8\%}}$ & $0.2\%_{\textcolor{Green}{-69.2\%}}$ \\
Qwen2.5-14B & oxy-1-small & Role play & 58.1\% & $74.7\%_{\textcolor{red}{+16.6\%}}$ & $0.2\%_{\textcolor{Green}{-57.9\%}}$ \\
Phi-4-mini & phi-4-mini-chinese-it-e1 & Reasoning \& STEM & 25.7\% & $58.0\%_{\textcolor{red}{+32.3\%}}$ & $1.4\%_{\textcolor{Green}{-24.3\%}}$ \\
Phi-4 & Meditron3-Phi4-14B & Clinical medicine & 25.2\% & $39.9\%_{\textcolor{red}{+14.7\%}}$ & $1.6\%_{\textcolor{Green}{-23.6\%}}$ \\
gemma-2b-it & gemma-2b-openhermes & Japanese language & 22.0\% & $26.3\%_{\textcolor{red}{+4.3\%}}$ & $0.8\%_{\textcolor{Green}{-21.2\%}}$ \\
gemma-7b-it & gemma-ko-7b-it-v0.40 & Korean language & 34.0\% & $38.6\%_{\textcolor{red}{+4.6\%}}$ & $1.0\%_{\textcolor{Green}{-33.0\%}}$ \\
Qwen3-4B & qwen3-4b-code-reasoning-v2 & Programming & 65.1\% & $70.7\%_{\textcolor{red}{+5.6\%}}$ & $0.5\%_{\textcolor{Green}{-64.6\%}}$ \\
Qwen3-14B & Fast-Math-Qwen3-14B & Math & 50.9\% & $61.1\%_{\textcolor{red}{+10.2\%}}$ & $1.1\%_{\textcolor{Green}{-49.8\%}}$ \\
\midrule
\multicolumn{3}{c|}{\emph{Average}} &
44.5\% &
$\textit{53.8\%}_{\textcolor{red}{\textit{+9.3\%}}}$ &
$\textit{0.8\%}_{\textcolor{Green}{\textit{-43.7\%}}}$\\
\bottomrule
\end{tabular}
\label{tab:posthoc_safety_hardening}
\end{table*}

\subsection{Post-Hoc Downstream Hardening}
We next evaluate whether the safety structure identified by \ourname{} can be reused to harden downstream fine-tuned variants of the same base model. This setting reflects a common deployment scenario: a provider releases or hosts a safety-aligned base model, after which downstream developers fine-tune it for specialized domains, languages, or reasoning tasks. Such downstream adaptation can alter refusal behavior and weaken safety alignment, even when the downstream task is not explicitly harmful. \ourname{} targets this problem by reusing the base model’s safety-neuron structure to restore safety post hoc, avoiding full safety realignment from scratch.

Table~\ref{tab:posthoc_safety_hardening} reports ASR before and after downstream fine-tuning, with and without post-hoc \ourname{} hardening. Downstream fine-tuning often increases vulnerability to jailbreaks. Across 10 downstream models, as expected, the average ASR rises from 44.5\% for the baseline models to 53.8\% after downstream fine-tuning. The degradation is especially pronounced for several specialized models: Phi-4-mini fine-tuned for Chinese reasoning and STEM increases from 25.7\% to 58.0\%, oxy-1-small increases from 58.1\% to 74.7\%. These results show that downstream adaptation can substantially erode safety robustness.

Applying post-hoc \ourname{} hardening consistently restores safety across all downstream targets. After hardening, the average ASR drops from 53.8\% to 0.8\%, corresponding to a 53.0\% reduction relative to the downstream models before hardening and a 43.7\% reduction relative to the original base-model baselines. The improvement is consistent across all model families and downstream tasks: \ourname{} reduces ASR to at most 1.6\% in every case.
These results demonstrate that the safety prior, i.e., neuron clusters, identified in the base model provide a reusable safety prior for downstream variants. Rather than rediscovering safety-relevant structure or updating broad parameter regions, \ourname{} carries over the selected neurons and cluster assignments from the base model and performs lightweight recovery training on the downstream model. This enables efficient post-hoc safety restoration after task-specific fine-tuning, supporting the maintainability goal of \ourname{} in settings where many downstream variants must be hardened over time.

\subsection{Utility Analysis}
We evaluate model utility before and after \ourname fine-tuning on three standard reasoning benchmarks: GSM8K~\cite{cobbe2021training} for multi-step mathematical reasoning, ARC~\cite{clark2018think} for abstract and commonsense reasoning through multiple-choice science questions, and MMLU~\cite{hendrycks2020measuring} for broad-domain knowledge and reasoning across diverse academic subjects. These benchmarks capture mathematical, logical, and general knowledge reasoning performance.

\begin{figure*}[htbp]
    \centering
    \subfloat[GSM8K]{%
        \includegraphics[width=0.325\textwidth]{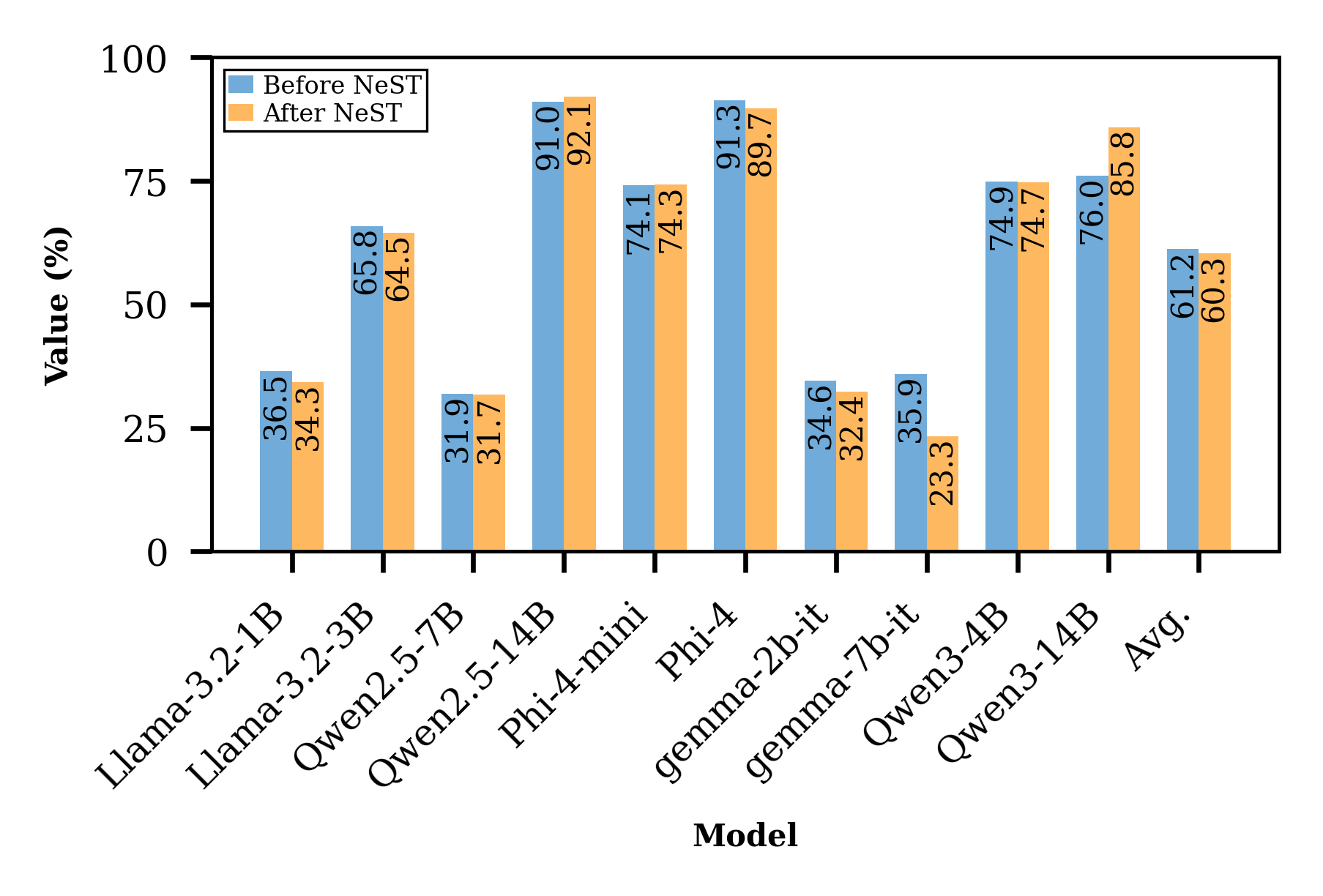}
    }
    \hfill
    \subfloat[ARC]{%
        \includegraphics[width=0.325\textwidth]{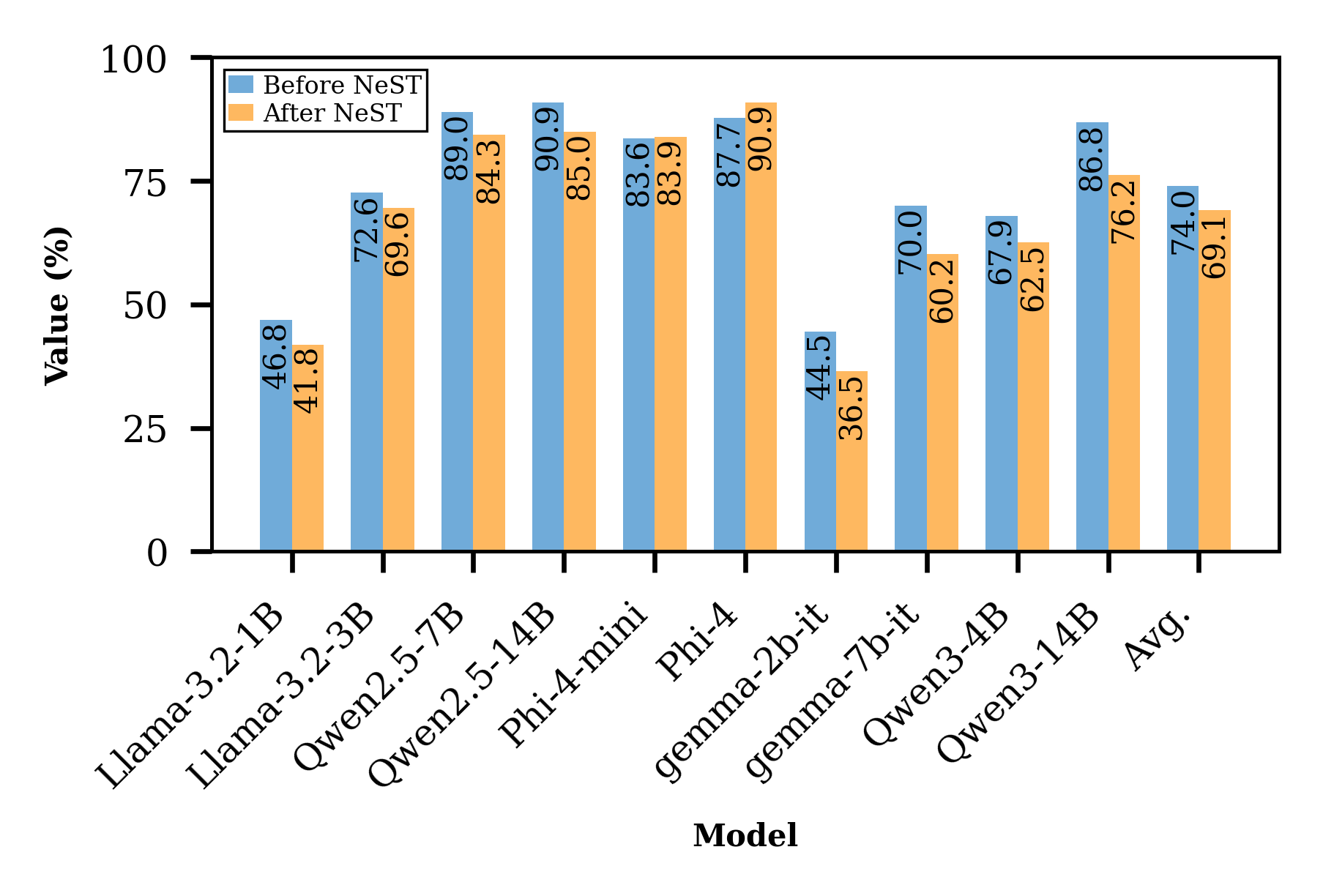}
    }
    \hfill
    \subfloat[MMLU]{%
        \includegraphics[width=0.325 \textwidth]{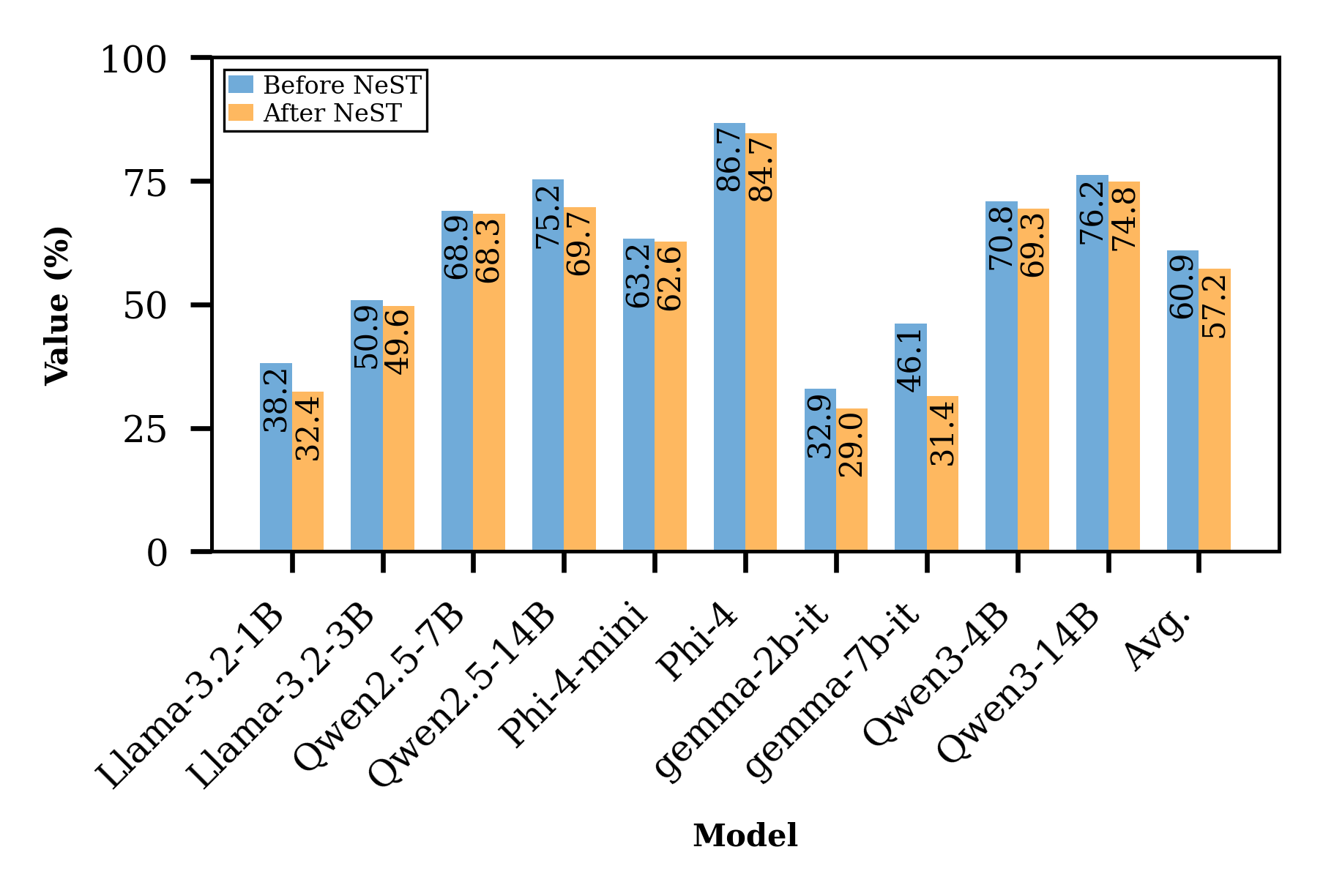}
    }

    \caption{Utility evaluation before and after \ourname{} fine-tuning across three reasoning benchmarks.}
    \label{fig:utility-analysis}
\end{figure*}

As shown in Figure~\ref{fig:utility-analysis},
\ourname{} preserves model utility across all benchmarks, with only minor performance changes. On GSM8K, the average accuracy drops by only 0.9 percentage points (from 61.2\% to 60.3\%). Similarly, ARC shows a modest average decrease from 74.0\% to 69.1\%, while MMLU exhibits a small drop from 60.9\% to 57.2\%.
Although some smaller models, such as gemma-2b-it, show more noticeable declines, several models, including Qwen2.5-14B, Qwen3-4B, and Phi-4, maintain or slightly improve performance. Indeed, GSM8K, ARC, and MMLU count refusals as incorrect answers; preserved (and sometimes improved) performance suggests NeST does not cause over-refusal.
These results indicate that \ourname{} preserves model capability across diverse reasoning tasks without substantially degrading utility.

%% file: Sections_submission/6_ablation.tex
\section{Ablation and Hyperparameter Study}
\label{sec:ablation study}

\subsection{Stability Across Jailbreaks and Judges}
\label{subsec:stability dataset_and_judge}
\begin{table*}[t]
\centering
\scriptsize
\caption{Attack success rate (ASR) across target models under two evaluation settings: JailBreakV-28k as the evaluation dataset and Llama-Guard-3-8B as the judge.}
\resizebox{\textwidth}{!}{
\begin{tabular}{l|c|c|c|c|c|c|c||c|c|c|c|c|c|c}
\toprule
\multirow{4}{*}{\textbf{Target Model}}
& \multicolumn{7}{c||}{\textbf{JailBreakV-28k Dataset}}
& \multicolumn{7}{c}{\textbf{Llama-Guard-3-8B Judge}} \\
\cmidrule(lr){2-8} \cmidrule(lr){9-15}
& \multirow{2}{*}{\textbf{Baseline}}
& \multirow{2}{*}{\textbf{Full FT}}
& \multicolumn{2}{c|}{\textbf{Circuit Breaker}}
& \multirow{2}{*}{\textbf{SN-Tune}}
& \multirow{2}{*}{\textbf{LoRA}}
& \multirow{2}{*}{\textbf{\ourname{}}}
& \multirow{2}{*}{\textbf{Baseline}}
& \multirow{2}{*}{\textbf{Full FT}}
& \multicolumn{2}{c|}{\textbf{Circuit Breaker}}
& \multirow{2}{*}{\textbf{SN-Tune}}
& \multirow{2}{*}{\textbf{LoRA}}
& \multirow{2}{*}{\textbf{\ourname{}}} \\
\cmidrule(lr){4-5} \cmidrule(lr){11-12}
& & & \textbf{Orig.} & \textbf{Exp.} & & &
& & & \textbf{Orig.} & \textbf{Exp.} & & & \\
\midrule
Llama-3.2-1B  & 23.1\% & 4.1\% & 32.2\% & 41.9\% & 1.1\%  & 9.7\% & 0.3\% & 15.1\% & 14.8\% & 18.6\% & 27.9\% & 0.1\%  & 12.6\% & 3.3\% \\
Llama-3.2-3B  & 38.0\% & 0.8\% & 38.9\% & 67.3\% & 11.0\% & 9.1\% & 1.5\% & 18.6\% & 4.4\%  & 11.1\% & 19.2\% & 4.0\%  & 3.4\%  & 1.7\% \\
Qwen2.5-7B    & 57.8\% & 1.6\% & 26.4\% & 63.2\% & 32.4\% & 0.8\% & 1.4\% & 28.7\% & 0.3\%  & 21.9\% & 31.8\% & 2.4\%  & 1.9\%  & 8.1\% \\
Qwen2.5-14B   & 26.7\% & 0.2\% & 12.8\% & 58.6\% & 31.8\% & 1.1\% & 1.3\% & 15.7\% & 0.0\%  & 11.3\% & 27.5\% & 6.4\%  & 0.95\% & 2.5\% \\
Phi-4-mini    & 48.6\% & 0.1\% & 24.9\% & 32.7\% & 21.4\% & 6.1\% & 4.3\% & 16.6\% & 1.4\%  & 14.2\% & 18.6\% & 13.1\% & 1.2\%  & 1.1\% \\
Phi-4         & 32.3\% & 0.7\% & 21.5\% & 77.0\% & 43.1\% & 1.4\% & 0.9\% & 11.2\% & 2.8\%  & 9.4\%  & 26.3\% & 18.0\% & 3.0\%  & 1.0\% \\
gemma-2b-it   & 9.1\%  & 0.2\% & 20.3\% & 28.9\% & 28.5\% & 2.7\% & 0.0\% & 7.7\%  & 3.4\%  & 10.3\% & 17.4\% & 13.5\% & 6.3\%  & 2.7\% \\
gemma-7b-it   & 45.7\% & 0.3\% & 31.6\% & 56.8\% & 44.6\% & 3.1\% & 2.2\% & 12.0\% & 4.2\%  & 15.4\% & 33.2\% & 18.4\% & 3.8\%  & 5.5\% \\
Qwen3-4B      & 59.8\% & 1.1\% & 50.4\% & 41.2\% & 63.2\% & 5.9\% & 1.3\% & 27.6\% & 0.1\%  & 19.2\% & 32.8\% & 22.8\% & 7.2\%  & 11.0\% \\
Qwen3-14B     & 45.5\% & 0.4\% & 22.3\% & 63.3\% & 55.0\% & 9.9\% & 1.3\% & 16.9\% & 0.1\%  & 9.7\%  & 24.8\% & 13.7\% & 3.2\%  & 1.6\% \\
\midrule
\emph{Average}
& \emph{38.7\%}
& \emph{1.0\%}
& \emph{28.1\%}
& \emph{53.1\%}
& \emph{33.2\%}
& \emph{5.0\%}
& \emph{1.5\%}
& \emph{17.0\%}
& \emph{3.2\%}
& \emph{14.1\%}
& \emph{26.0\%}
& \emph{11.2\%}
& \emph{4.4\%}
& \emph{3.9\%} \\
\bottomrule
\end{tabular}
}
\label{tab:asr_merged}
\end{table*}

To verify that our conclusions are not specific to \texttt{WildJailbreak}, we additionally evaluate on \texttt{JailBreakV-28k}~\cite{luo2024jailbreakv28k} with 1,000 jailbreak prompts. As shown in Table~\ref{tab:asr_merged} (JailBreakV-28k column), the overall trends remain consistent on JailBreakV-28k, indicating that the results are not specific to a single benchmark. Full FT achieves the lowest average ASR at 1.0\%, while Circuit Breaker again shows substantially worse performance under our unified evaluation setting, increasing from 28.1\% to 53.1\% average ASR. SN-Tune reduces ASR relative to the baseline but still exhibits a relatively high average ASR of 33.2\%. LoRA achieves 5.4\% average ASR, while \ourname{} further reduces this to 1.5\% using lightweight neuron-level updates. Overall, the results closely mirror those observed on WildJailbreak, suggesting that \ourname{} generalizes stably across different jailbreak benchmarks.

We have also evaluated the results with a different judge, i.e., Llama-Guard-3-8B~\cite{grattafiori2024llama}. The results are shown in Table~\ref{tab:asr_merged} (Llama-Guard-3-8B column). 
Compared to the Qwen3Guard evaluation, all methods generally obtain lower ASR values under Llama-Guard-3-8B, suggesting a stricter or more conservative judging behavior. Nevertheless, the relative ordering of methods remains stable. Full FT continues to achieve the lowest ASR overall, while Circuit Breaker exhibits a substantial increase under the expanded evaluation setting, rising from 14.1\% to 26.0\% average ASR. SN-Tune also maintains noticeably higher ASR than both LoRA and \ourname{}, with an average ASR of 11.2\%. Similar to the main results, LoRA and \ourname{} remain the strongest lightweight approaches, achieving average ASRs of 4.4\% and 3.9\%, respectively. Importantly, \ourname{} preserves low ASR across multiple model families, including Phi, Gemma, Qwen, and Llama models, further supporting the stability and robustness of the proposed neuron-level hardening approach across different judge models. 

\subsection{Over-Refusal Analysis}
\label{subsec:overrefusal}

While reducing harmful generations is important, an effective safety alignment method should also avoid unnecessarily refusing benign requests. We therefore evaluate the over-refusal rate (ORR) of \ourname{} using 500 benign samples from \texttt{Natural Reasoning}~\cite{yuan2025naturalreasoningreasoningwild28m}. Lower ORR indicates better preservation of helpful behavior.

As shown in the ORR column of Table~\ref{tab:jaccard_stability}, \ourname{} achieves a low average ORR of 0.9\% across all evaluated models, with consistently low refusal rates across different model families. This suggests that the substantial reduction in ASR is not achieved by broadly suppressing model outputs. Instead, \ourname{} selectively targets safety-relevant behaviors while preserving the ability to answer benign instructions, indicating a favorable balance between safety and utility.
\begin{table}[t]
\centering
\scriptsize
\caption{Over-refusal rate (ORR) of NeST on benign prompts and Jaccard overlap of selected safety neurons across different probe seeds and harmful datasets.}
\begin{tabular}{l|c||c|c}
\toprule
\textbf{Target Model} & \textbf{ORR} & \textbf{Seed Jaccard} & \textbf{Dataset Jaccard} \\
\midrule
Llama-3.2-1B  & 1.6\% & 90.5\% & 90.8\% \\
Llama-3.2-3B  & 1.2\% & 87.6\% & 94.9\% \\
Qwen2.5-7B    & 0.2\% & 63.3\% & 83.9\% \\
Qwen2.5-14B   & 0.3\% & 67.4\% & 93.3\% \\
Phi-4-mini    & 1.2\% & 73.5\% & 89.3\% \\
Phi-4                  & 0.8\% & 68.4\% & 91.7\% \\
gemma-2b-it            & 1.2\% & 78.9\% & 88.7\% \\
gemma-7b-it            & 0.9\% & 85.9\% & 84.3\% \\
Qwen3-4B               & 0.7\% & 69.4\% & 93.7\% \\
Qwen3-14B              & 0.5\% & 83.1\% & 91.5\% \\
\midrule
\emph{Average} & \emph{0.9\%} & \emph{76.8\%} & \emph{90.2\%} \\
\bottomrule
\end{tabular}
\label{tab:jaccard_stability}
\end{table}

\subsection{Stability of Safety-Neuron Detection} 
\label{subsec:Stability of Safety-Neuron Detection}
We evaluate the stability of safety-neuron identification using Jaccard similarity analysis under two settings: different probe-training seeds and different harmful datasets. The goal is to determine whether the identified safety neurons remain consistent across varying experimental conditions.
In the first experiment, we vary the random seed used for probe training while keeping the model, datasets, activation extraction procedure, and neuron-selection threshold fixed. Safety neurons are identified independently for each run using the same logistic regression probing framework, and the overlap between selected neuron sets is measured by Jaccard similarity. High overlap indicates that the neuron-selection process is stable despite training randomness.
In the second experiment, we evaluate robustness across different harmful datasets while keeping the benign prompts and probing configuration fixed. We compare neuron sets identified using \texttt{walledai/CatHarmfulQA} and \texttt{LLM-LAT/harmful-dataset}. Safety neurons are extracted independently for each dataset pair, and both layer-wise and global Jaccard overlap scores are computed.

Table~\ref{tab:jaccard_stability} shows that safety-neuron selection is stable under both perturbations. Across all models, Seed Jaccard averages 76.8\%, indicating substantial overlap despite stochastic probe training and the conservativeness of exact-set comparison. Llama models exhibit particularly high seed stability, with overlaps of 90.5\% and 87.6\%. Qwen2.5 and Phi models show lower Seed Jaccard, ranging from 63.3\% to 73.5\%, but these values still indicate that most selected neurons are shared across runs. The lower seed overlap mainly reflects variation among near-threshold or redundant neurons, rather than a complete change in the detected safety structure.
Dataset Jaccard is even higher, averaging 90.2\%. This indicates that the selected safety neurons are largely preserved when the harmful prompt source changes. The overlap exceeds 90\% for most models. These results suggest that \ourname{} identifies safety-relevant neurons associated with general harmfulness recognition and refusal behavior, rather than neurons specific to a single probing dataset.
Overall, the Jaccard analysis shows that \ourname{} produces a stable safety-neuron prior. Seed variation mainly affects borderline neurons under an exact thresholded set metric, while the high dataset overlap demonstrates that the core safety-neuron structure generalizes across harmful prompt sources.

\subsection{Impact of the z-Threshold}
\label{subec:Impact of the z-Threshold}
The z-threshold ($z_{\text{thr}}$) controls the selectivity of safety neuron detection by specifying how strongly a neuron’s probe score must deviate from the layer-wise mean to be considered safety-relevant, as formalized in Eq.~\eqref{eq:safety_neurons}. Lower thresholds include a larger set of neurons, potentially introducing noisy or weakly relevant features, while higher thresholds yield smaller and more selective neuron sets that may under-represent safety-critical behavior.

We evaluate three representative thresholds, $z_{\text{thr}} \in {2,3,4}$, while keeping all other components of \ourname{} fixed. Table~\ref{tab:abl_z_thres} reports the resulting Attack Success Rates (ASR) across ten target models. On average, the lowest threshold, $z_{\text{thr}}=2$, achieves the best overall performance with an average ASR of 0.54\%, followed by $z_{\text{thr}}=3$ with an average ASR of 1.1\%. In contrast, the more restrictive setting $z_{\text{thr}}=4$ substantially degrades performance, increasing the average ASR to 3.7\%. This nearly 7$\times$ increase relative to $z_{\text{thr}}=2$ suggests that overly aggressive neuron filtering removes safety-critical neurons and weakens safety adaptation. Overall, the results indicate that retaining a broader set of safety-relevant neurons yields more robust safety performance across model families.

\begin{table}[ht]
\centering
\scriptsize
\caption{Impact of the z-threshold.}
\begin{tabular}{l|ccc}
\toprule
\textbf{Target Model} & \textbf{$z_{\text{thr}}=2$} & \textbf{$z_{\text{thr}}=3$} & \textbf{$z_{\text{thr}}=4$}\\
\midrule
Llama-3.2-1B & 0.0\% & 0.3\% & 1.9\% \\
Llama-3.2-3B & 0.0\% & 0.5\% & 0.3\% \\
Qwen2.5-7B & 0.2\% & 0.6\% & 4.7\% \\
Qwen2.5-14B & 0.1\% & 0.8\% & 2.3\% \\
Phi-4-mini & 0.8\% & 0.7\% & 5.3\% \\
Phi-4 & 0.0\% & 1.3\% & 0.9\% \\
gemma-2b-it & 1.1\% & 0.2\% & 5.1\% \\
gemma-7b-it & 2.0\% & 5.3\% & 9.2\% \\
Qwen3-4B & 0.1\% & 0.0\% & 2.8\% \\
Qwen3-14B & 1.1\% & 1.0\% & 4.9\% \\
\midrule
\emph{Average} & \emph{0.54\%} & \emph{1.1\%} & \emph{3.7\%} \\
\bottomrule
\end{tabular}
\label{tab:abl_z_thres}
\end{table}

Beyond the average case, the results show that NeST is relatively robust to moderate threshold variations. Both $z_{\text{thr}}=2$ and $z_{\text{thr}}=3$ achieve consistently low ASR across most models, with average ASRs of 0.54\% and 1.1\%, respectively.

In contrast, increasing the threshold to $z_{\text{thr}}=4$ consistently degrades performance, raising the average ASR to 3.7\%. This effect is particularly evident for models such as Phi-4-mini, gemma-2b-it, and Qwen2.5-7B, which exhibit substantial ASR increases under the stricter threshold. These results suggest that overly aggressive neuron filtering removes a meaningful fraction of safety-critical neurons, thereby reducing the effectiveness of subsequent safety adaptation.

\subsection{Impact of Clustering Strength}
\label{subsec:Impact of Clustering Strength}
In this section, we control the degree of parameter sharing among safety neurons through the silhouette threshold used during clustering. Three settings are considered: (i) weak clustering (minimum setting) allocates only a single cluster per layer, meaning that all safety neurons within a layer share the same adaptation parameters; 
(ii) strong clustering (maximum setting) treats each safety neuron as its own cluster, thus removing parameter sharing and allowing fully independent adaptation;
(iii) the default \ourname{} setting lies between these two extremes, grouping safety neurons into a moderate number of clusters per layer based on activation similarity.

\begin{table}[ht]
\centering
\scriptsize
\caption{Ablation of safety-neuron selection strength.}
\begin{tabular}{l|cc|cc|cc}
\toprule
\multirow{2}{*}{\textbf{Target Model}}
& \multicolumn{2}{c|}{\textbf{Weak}}
& \multicolumn{2}{c|}{\textbf{\ourname{}}}
& \multicolumn{2}{c}{\textbf{Strong}} \\
\cmidrule(lr){2-3}
\cmidrule(lr){4-5}
\cmidrule(lr){6-7}
& \textbf{ASR} & \textbf{Para.}
& \textbf{ASR} & \textbf{Para.}
& \textbf{ASR} & \textbf{Para.} \\
\midrule
Llama-3.2-1B & 1.2\% & 0.1 & 0.3\% & 0.1 & 0.7\% & 1.4 \\
Llama-3.2-3B & 1.1\% & 0.2 & 0.5\% & 0.3 & 0.0\% & 4.2 \\
Qwen2.5-7B & 0.1\% & 0.2 & 0.6\% & 0.4 & 0.3\% & 5.7 \\
Qwen2.5-14B & 1.1\% & 0.4 & 0.8\% & 1.0 & 0.1\% & 12.1 \\
Phi-4-mini & 3.4\% & 0.1 & 0.7\% & 0.2 & 0.4\% & 4.2 \\
Phi-4 & 0.9\% & 0.2 & 1.3\% & 0.4 & 0.0\% & 10.3 \\
gemma-2b-it & 2.5\% & 0.1 & 0.2\% & 0.2 & 0.6\% & 0.5 \\
gemma-7b-it & 2.9\% & 1.7 & 5.3\% & 0.4 & 0.7\% & 3.6 \\
Qwen3-4B & 0.9\% & 0.1 & 0.0\% & 0.4 & 0.1\% & 2.5 \\
Qwen3-14B & 0.7\% & 0.4 & 1.0\% & 0.9 & 0.5\% & 8.6 \\
\midrule
\emph{Average}
& \emph{1.5\%} & \emph{0.3}
& \emph{1.1\%} & \emph{0.4}
& \emph{0.3\%} & \emph{5.3} \\
\bottomrule
\end{tabular}
\label{tab:abl_clustering}
\end{table}

Table~\ref{tab:abl_clustering} reports the resulting attack success rates across models. On average, weak clustering yields an ASR of 1.5\%, while the default \ourname{} setting reduces it to 1.1\%. Strong clustering further lowers the average ASR to 0.3\%, achieving the best overall safety performance. However, this improvement comes at a substantially higher parameter cost. While the default \ourname{} configuration requires only 0.4M trainable parameters on average, strong clustering increases this cost to 5.3M parameters, representing more than a 12$\times$ increase in trainable parameters for a reduction of only 0.8 percentage points in ASR. These results highlight the trade-off between parameter efficiency and adaptation flexibility. Under weak clustering, all safety neurons within a layer share a highly constrained set of updates, resulting in the lowest parameter cost (0.3M on average) but also the highest ASR. This suggests that overly aggressive parameter sharing limits the model's ability to capture diverse safety-relevant behaviors, leading to underfitting.

In contrast, strong clustering maximizes adaptation flexibility by allowing more fine-grained updates to safety neurons. This provides the highest degree of adaptation capacity and yields the lowest average ASR across the evaluated models. Strong clustering improves performance on several models, e.g., Llama-3.2-3B, Qwen2.5-7B, Qwen2.5-14B, Phi-4-mini, Phi-4, and gemma-7b-it. However, these gains come at a substantially increased parameter cost. Moreover, in certain models, including Llama-3.2-1B, gemma-2b-it and Qwen3-4B, strong clustering even degrades performance relative to the default setting, despite using significantly more trainable parameters. This suggests that fully independent neuron-level updates can introduce instability or overfitting, indicating that moderate parameter sharing help improve robustness while maintaining strong safety performance.

Overall, these results highlight a clear efficiency-performance trade-off. Weak clustering minimizes computation and memory overhead but sacrifices safety effectiveness, while strong clustering substantially increases adaptation cost for relatively modest average gains. This suggests that the default \ourname{} configuration already captures much of the safety-relevant neuron subspace while avoiding the overhead and potential instability associated with fully independent updates. As a result, the default setting balances between safety performance and parameter efficiency.

%% file: Sections_submission/7_related_works.tex
\section{Discussion}
\label{sec:discussion}

\noindent\textbf{Privileged Attacks and Defenses.}
As defined in Section~\ref{subsec:threat model}, \ourname{} is designed for hosted black-box deployment settings. This setting reflects the main use case of \ourname{}: strengthening the model's intrinsic refusal behavior before deployment, without adding inference-time mechanisms or exposing model internals to users. White-box attacks that inspect, optimize against, or directly manipulate internal representations~\cite{wu2025neurostrike,krauss2025twinbreak} require a substantially more privileged adversary than the one considered in this work. Such attacks are important for open-weight release scenarios, but they evaluate a different security problem from hosted black-box safety hardening. Indeed, under unrestricted white-box access, an adversary can operate directly on the final model, its gradients, and its parameters, regardless of whether the safety behavior was introduced by supervised safety tuning, RLHF, DPO, LoRA-based adaptation, or \ourname{}. When the hosted deployment assumption is preserved through secure model hosting, access control, or similar protections, \ourname{} provides a strong and lightweight mechanism for strengthening the model's intrinsic safety.

Within our black-box threat model, adaptive attacks remain relevant when they operate only through model queries. Gradient-based optimization methods such as Greedy Coordinate Gradient (GCG)~\cite{zou2023universal,liao2024amplegcg,xu2026routehijack} are not directly applicable because they require access to model gradients to compute adversarial suffixes. Mutation-based black-box methods such as LLM-Fuzzer~\cite{yu2024llm} instead rely on strong initial jailbreak seeds and output-oracle feedback to guide the search. In our experiments, this feedback becomes sparse once the model rejects most seed prompts, making adaptive mutation substantially less effective. Empirically, running LLM-Fuzzer on Phi-4 with its default seed corpus yields a 0.78\% ASR over 20,000 queries under the same Qwen3Guard-based evaluation protocol used in Section~\ref{sec:Performance Evaluation}. After applying \ourname{}, the ASR is further reduced to 0.03\%.

\noindent\textbf{Task-Specific and Structure-Aware Adaptation.}
Unlike LoRA, which is primarily designed as a lightweight, general-purpose adaptation mechanism, \ourname{} is task-structured by design. While LoRA introduces low-rank updates without regard to the semantic role of individual neurons, \ourname{} explicitly aligns adaptation with the internal structure of a target behavior. In this work, we instantiate \ourname{} for safety alignment by detecting and adapting safety-relevant neurons, but the framework itself is not inherently limited to safety. The central design principle of \ourname{} is to first identify neurons that are causally and functionally associated with a given behavior, and then restrict adaptation to those neurons in a structured manner. This formulation naturally extends beyond safety alignment. By modifying the neuron detection procedure, \ourname{} can be applied to harden or specialize models for other targeted behaviors. In such settings, neuron detection is performed with respect to the relevant behavioral distinction, while the same cluster-based adaptation mechanism enables lightweight, behavior-specific fine-tuning aligned with the model’s internal structure. Viewed more broadly, \ourname{} provides a general framework for neuron-structured, task-specific parameter-efficient adaptation. Unlike generic parameter-efficient methods that apply uniform constraints across layers or projections, \ourname{} defines where to adapt based on internal representation structure, making it particularly well-suited for tasks where the target behavior is within specific neuron groups.

\noindent\textbf{Compatibility with Diverse Safety Alignment Methods.}
In this work, \ourname{} is implemented on top of supervised fine-tuning for safety alignment; however, the framework itself is not tied to a specific optimization objective. The neuron detection, clustering, and selective adaptation mechanisms operate directly at the level of model parameters and gradients, making \ourname{} compatible with a wide range of alignment paradigms. In principle, \ourname{} can be integrated with reinforcement learning-based approaches such as reinforcement learning from human feedback (RLHF), group-relative policy optimization (GRPO), and other preference-based optimization methods. In these settings, \ourname{} would restrict gradient updates during policy optimization to the clustered safety-relevant neurons, rather than allowing unrestricted parameter updates. This positions \ourname{} as a structural constraint on alignment optimization, agnostic to whether the underlying objective is supervised, preference-based, or reinforcement-driven.

\section{Related Work}
\label{sec:related}

Practical safety adaptation for LLMs requires both efficiency and reliability. A safety update should be lightweight enough to apply repeatedly across models and deployment settings, while still targeting the internal mechanisms that govern refusal behavior. Existing work approaches this problem from two main directions. Parameter-efficient fine-tuning reduces the cost of model adaptation, whereas safety-localization and intervention methods study or manipulate internal components associated with unsafe generation. \ourname{} bridges these directions by making safety adaptation both parameter-efficient and structure-aware.

\noindent\textbf{Parameter-efficient fine-tuning.}
Parameter-efficient fine-tuning (PEFT) methods adapt LLMs by updating only a small fraction of parameters. Representative approaches include adapter-based tuning~\cite{houlsby2019parameter,pfeiffer2020adapterhub}, bias-only tuning~\cite{zaken2022bitfit}, scaling-based tuning~\cite{liu2022few}, and LoRA~\cite{hu2022lora}. LoRA has become especially widely used because it freezes pretrained weights and injects trainable low-rank updates into existing linear transformations. These methods substantially reduce training and storage costs, but their update spaces are typically defined at generic architectural units such as layers, projections, or adapters. As a result, they do not explicitly account for where safety behavior is represented inside the network. In contrast, \ourname{} defines the update space using safety-relevant internal structure: it identifies neurons associated with refusal behavior, groups them into activation-based clusters, and trains only shared cluster-level updates.

\noindent\textbf{Safety localization and neuron-level tuning.}
A growing body of work studies safety behavior through internal representations, motivated by evidence that refusal and harmfulness-related behavior can be localized to specific layers, activations, or neurons~\cite{mu2020compositional,antverg2021pitfalls,wu2025neurostrike,wu2025gatebreaker}. These studies suggest that safety is not only an output-level property, but is also reflected in the model's internal computation. SN-Tune~\cite{zhao2025understanding} further shows that selectively modifying safety-relevant neurons can improve refusal behavior while leaving most model parameters unchanged. However, such neuron-level methods typically identify safety neurons by measuring their importance for harmful-query processing, for example through deactivation or perturbation-based analysis. This focuses primarily on sensitivity to harmful prompts and does not explicitly contrast safety-critical activations against benign behavior. Moreover, detected neurons are often tuned directly, without modeling functional structure among them. \ourname{} differs in two ways. First, it uses contrastive activation probing over harmful and benign prompts to identify neurons that distinguish safety-critical contexts from benign ones. Second, it clusters these neurons by activation behavior and learns shared updates within each cluster, yielding a more structured and coherent neuron-level safety adaptation mechanism.

\noindent\textbf{Representation-level safety interventions.}
Another line of work mitigates unsafe behavior by intervening on internal representations. Circuit Breakers~\cite{zou2406improving}, for example, learn auxiliary LoRA-based modules that reroute harmful activation trajectories toward refusal or incoherent states. Such methods can reduce unsafe outputs without full-parameter fine-tuning, but they introduce additional learned mechanisms that act as representation-level controls. Their effectiveness can also depend on how well the training data captures the harmful representation patterns encountered at test time. \ourname{} instead performs training-time safety hardening through mergeable neuron-level updates. After learning cluster-level updates, \ourname{} folds them into the original model weights, producing a standard model with no additional inference-time overhead.

Overall, prior approaches either provide parameter-efficient adaptation without modeling safety-specific internal structure, or exploit safety-related internal structure without offering a compact, cluster-based update mechanism. \ourname{} combines these advantages by localizing safety-relevant neurons, organizing them into functionally coherent clusters, and training shared updates only within those clusters. The resulting safety prior, consisting of the selected safety neurons and their cluster assignments, can be reused for both base-model hardening and post-hoc hardening of downstream fine-tuned variants.

%% file: Sections_submission/8_conclusion.tex
\section{Conclusions}
\label{sec:conclusions}

This paper presents \ourname{}, a lightweight and structure-aware framework for post-hoc safety hardening of large language models. Rather than updating the full model or attaching generic adapters, \ourname{} identifies safety-relevant neurons through contrastive activation probing, organizes them into functionally coherent clusters, and trains shared cluster-level updates while freezing the rest of the model. The learned updates are then folded into the original weights, yielding a standard hardened model with no additional inference-time cost.
Across 14 open-weight language and multimodal models, \ourname{} substantially reduces attack success rates while using orders-of-magnitude fewer trainable parameters than full fine-tuning and fewer parameters than existing lightweight baselines. It also preserves core reasoning and knowledge capabilities, avoids excessive over-refusal, and remains effective across different jailbreak benchmarks, judges, input modalities, and inference settings. Beyond hardening base models, \ourname{} provides a reusable safety prior: the selected safety neurons and cluster assignments can be transferred to downstream fine-tuned variants, enabling efficient post-hoc safety restoration without rediscovering safety structure or repeating full safety alignment.
\ourname{} shows that robust safety alignment can be achieved through localized, cluster-level adaptation, offering a practical path for maintaining LLM safety as models are repeatedly fine-tuned, specialized, and deployed.

%% file: Sections_submission/10_appendix.tex
\newpage
\appendices

\section{Ethical Considerations}
\textbf{Overview.}
This work proposes \ourname{}, a lightweight and structure-aware framework for strengthening safety alignment in large language models. The objective is to reduce unsafe generations, improve robustness to harmful prompts, and support safer deployment of open-weight and multimodal LLM systems.

\textbf{Stakeholders.}
Relevant stakeholders include LLM developers and deployers, organizations integrating AI into user-facing systems, safety researchers, and end users who may be affected by unsafe model behavior. The method is intended to improve reliability while preserving useful model capabilities.

\textbf{Data and Experimental Protocols.}
All experiments use publicly available datasets, open-weight pretrained models, and controlled offline evaluation. No private, proprietary, or personally identifiable data were collected or processed, and no real-world deployment was involved.

\textbf{Potential Impact.}
\ourname{} strengthens intrinsic safety mechanisms while maintaining utility and enabling efficient post-hoc safety hardening after downstream fine-tuning, contributing to scalable and maintainable safety practices for evolving LLM systems.

\textbf{Mitigations for Negative Impacts.}
The framework operates entirely through offline training and evaluation and does not introduce mechanisms for generating harmful content or weakening safeguards. Released artifacts focus on defensive safety alignment and include documentation emphasizing responsible research use.

\textbf{Decision to Conduct and Publish the Study.}
We report this work to advance scalable safety alignment and support the development of more robust and trustworthy LLM deployments as real-world use continues to expand.

\textbf{Research Team Considerations.}
The study involves only controlled benchmark evaluation and follows standard responsible research practices to ensure safe handling of safety-related data and contributor well-being.